\documentclass{article} 
\usepackage{enumerate}     
\title{\bf     On some invariants in numerical semigroups and
estimations of the order bound.}  
\author{Anna Oneto  \ and \  Grazia Tamone }  
    
\date{} 

\expandafter\chardef\csname pre amssym.def  at\endcsname=\the\catcode`\@
\catcode`\@=11
 
\def\undefine#1{\let#1\undefined} 
\def\newsymbol#1#2#3#4#5{\let\next@\relax
 \ifnum#2=\@ne\let\next@\msafam@\else
 \ifnum#2=\tw@\let\next@\msbfam@\fi\fi
 \mathchardef#1="#3\next@#4#5}
\def\mathhexbox@#1#2#3{\relax
 \ifmmode\mathpalette{}{\m@th\mathchar"#1#2#3}%
 \else\leavevmode\hbox{$\m@th\mathchar"#1#2#3$}\fi}
\def\hexnumber@#1{\ifcase#1 0\or 1\or 2\or 3\or 4\or 5\or 6\or 7\or 8\or
 9\or A\or B\or C\or D\or E\or F\fi}
 
\font\tenmsa=msam10
\font\sevenmsa=msam7  
\font\fivemsa=msam5
\newfam\msafam
\textfont\msafam=\tenmsa
\scriptfont\msafam=\sevenmsa
\scriptscriptfont\msafam=\fivemsa 
\edef\msafam@{\hexnumber@\msafam}

\font\tenmsb=msbm10
\font\sevenmsb=msbm7
\font\fivemsb=msbm5
\newfam\msbfam
\textfont\msbfam=\tenmsb
\scriptfont\msbfam=\sevenmsb
\scriptscriptfont\msbfam=\fivemsb
\edef\msbfam@{\hexnumber@\msbfam}

   \font\tengothic=eufm10
   \font\sevengothic=eufm7
   \newfam\gothicfam
   \textfont\gothicfam=\tengothic
   \scriptfont\gothicfam=\sevengothic
   
   \font\tenmsb=msbm10
   \font\sevenmsb=msbm7
   \newfam\msbfam
   \textfont\msbfam=\tenmsb
   \scriptfont\msbfam=\sevenmsb
   						
\newtheorem{prop}{Proposition}[section] 
\newtheorem{set}[prop]{Setting} 
\newtheorem{rem}[prop]{Remark}
\newtheorem{thm}[prop]{Theorem}
\newtheorem{coro}[prop]{Corollary}
\newtheorem{defin}[prop]{Definition}

\newtheorem{lemma}[prop]{Lemma}
\newtheorem{notat}[prop]{Notation}
\newtheorem{ex}[prop]{Example}
\newtheorem{conj}[prop]{Conjecture}
\newtheorem{(*)}[prop]{}

\newcommand{\s}{\mbox{$\widetilde s$}}  
\newcommand{\n}{\mbox{$\#$}}

\newcommand{\e}{\mbox{\sf  {$\ell$}}}

\newcommand{\integ}{\mbox{${\sf Z\hspace{-1.3 mm}Z}$}}

\newcommand{\nat}{\mbox{${\rm I\hspace{-.6 mm}N}$}}
\newcommand{\natsmall}{\mbox{${\rm\scriptstyle I\hspace{-.6 mm}N}$}} 
\newcommand{\I}{\mbox{$\  \Longrightarrow \ $}}
\newcommand{\II}{\mbox{$\  \Longleftrightarrow \ $}}

\newcommand{\enu} {\begin{enumerate}[{\rm(1)}]} 
\newcommand{\enua} {\begin{enumerate}[ $(a)$]} 
\newcommand{\enui} {\begin{enumerate}[ $(i)$]} 
\newcommand{\denu} {\end{enumerate}}
\newcommand{\spa}{\vspace{0.2cm}}
\newcommand{\0} {\raisebox{.5mm}{$ \scriptstyle{\bigcirc }$} }

\font\tengothic=eufm10 
\font\sevengothic=eufm7
\newfam\gothicfam 
\textfont\gothicfam=\tengothic
\scriptfont\gothicfam=\sevengothic

\begin{document}
\maketitle	{\footnote {  
 \quad The first author is with  Diptem, 
  Universit\`a di Genova, \  P.le Kennedy,\newline Pad. D - 16129 Genova (Italy)
  ({\em E-mail}:   oneto@diptem.unige.it).
  The second author is with Dima,  Universit\`a di Genova,\newline  Via Dodecaneso 35 - 16146 Genova (Italy)
  ({\em E-mail}:   tamone@dima.unige.it).}} 
 {\bf    Abstract.}{ \small \   Let $S=\{s_i\}_{i\in\natsmall}\subseteq \nat$  be a  numerical semigroup.  For $s_i\in S$, let 
$\nu(s _ i)$   denote the number of pairs   $ (s_i\!-\!s_j,s_j)\in S^2 $.  When $S$ is the   Weierstrass semigroup  of a family
$\{{\cal C}_i\}_{i\in\natsmall}$ of one-point  algebraic-geometric  codes, a good bound for the   minimum distance   of the code ${\cal C}_i$ is the  Feng and Rao
 {\it order bound}  
 $d_{ORD}( C_i):= min \{\nu(s _j)  :\ j\geq i+1\}  $.  It is well-known that there exists an integer $m$ such that the  sequence $\{\nu(s _i)\}_{i\in\natsmall}$ is non-decreasing for  $  s_i> s_m$, therefore  $d_{ORD}( C_i) =\nu(s_{i+1})$ for \ $i\geq m$.  By way of some suitable parameters related to the semigroup $S$, we   find upper bounds for $s_m$, we   evaluate $s_m$ exactly in many cases, further we give a lower bound for several classes of semigroups.
 \vspace{-0.2cm} \\ 

{\em Index Therms.} Numerical semigroup, Weierstrass semigroup,  AG code, order bound on the minimum distance, Cohen-Macaulay type.}
 \section{Introduction}

  Let  $S=\{s_i\}_{i\in\natsmall} \subseteq \nat$    be a  numerical semigroup      and let 
  $ e, \ c,\ c  ',\ d,\ d  '\  $ denote respectively  the multiplicity, the conductor,
the subconductor,   the dominant of the semigroup and    the  greatest\ element    in $ S $ 
preceding $\ c  '$  (if $e>1$),  as    in Setting \ref{set1}. Further let   $  \ \ell\ $ be the number of gaps of $S$ between $d$ and $c$,    and let 
 \centerline{$\s :=max\{s\in S \  \ such \ that\  \ s\leq d\ \ and \ \ s-\e\notin S\} $.\qquad}  When  $S$ is the   Weierstrass semigroup  of a family
$\{{\cal C}_i\}_{i\in\natsmall}$ of one-point  $AG$ codes (see   \cite{hlp},\cite{fr1}), a good bound for the minimum distance of     ${\cal C}_i$ is the Feng and Rao 
 {\em order  bound} \\
\centerline{ 
 $d_{ORD}( C_i):= min \{\nu(s _j)  :\ j\geq i+1\}  $}
 where,     for   $s_j\in S$,  
$\nu(s _ j) $ denotes   the number of pairs   $ (s_j\!-\!s_k,s_k)\in S^2 $  (see    \cite{fr1}).
It is well-known that there exists an integer $m$ such that   sequence $\{\nu(s _i)\}_{i\in\natsmall}$ is non-decreasing for   $i\geq m+1$ (see\cite{kp}) and so   $d_{ORD}( C_i) =\nu(s_{i+1})$ for \ $i\geq m$.  For this reason it is important to find the element $s_m$ of $S$. 
In our papers \cite{ot1} and \cite{ot2}, we proved that $s_m=\s+d$ \  if \ $\s\geq d'$, further we evaluated $s_m$ for $\e\leq 2,  \ e\leq 6$,   \ $Cohen$-$Macaulay$ $type\leq 3$. \par   
In this paper, by a more detailed study  of the semigroup  we find interesting relations among the integers defined above; further   by using these relations  we  deduce the Feng and Rao 
  order bound in several new situations.  Moreover in every considered case we show  that  $s_m\geq c+d-e$. \\  In  Section 2, we   establish    various formulas and inequalities  among the integers $e,\ \e,\  d',\ c',\ d,\ c  $  and $t:=d-\s $,  see in particular (\ref{minor2d}) and (\ref{rec0}). In Section 3, by using the results of Section 2 and some result from \cite{ot2},  we improve the  known facts on $s_m$ recalled above; further we   state   the conjecture  that $c+d-e$ ia always a lower bound for $s_m$ and we prove it in many cases. Finally (Section 4) we deduce some particular case by applying the previous results and by some direct trick.\par
In conclusion by glueing togheter some facts of \cite{ba}, \cite{ot1}, \cite{ot2} and the results of the present paper,  we see that   the value of the {\it order bound} $s_m$  depends   essentially on the position of the integer $\s$ \ in the semigroup.  We   summarize   below the main results for the convenience of the reader.   $$\begin{array}{llllc}
{\rm If}& \s\geq d'+c'-d&{\rm then}&s_m=  \s+d \qquad   \\
{\rm if}& \s= 2d' -d&{\rm then} &s_m=  \s+d=2d' \qquad   \\
{\rm if}& \s\leq d'+c'-d-1&{\rm then} &s_m\leq  max\{\s+d, 2d'\}.\qquad  (\ref{all})\end{array}$$
 
$$\begin{array}{llllc}{\rm If} &   [d  '-\e,d  ']\cap\nat\subseteq
S&{\rm then}  & \left [\begin{array}{llll} \!\! \s+d  '-\e +1\leq s_m\leq 2d  '  &if\ \   2d  '-d <\s< d  '+c  '-d\\
 s_m=\s+d& {\it otherwise}.  \qquad (\ref{interv})\end{array}\right.\\
 \\
 {\rm If}&\left\{\begin{array}{lll}\s\leq d'-2\\
  \left [\begin{array}{l}\s +2, d'\end{array}\right] \cap\nat\subseteq S\\
   2d  '-d <\s< d  '+c  '-d\\
  \end{array}\right.  &{\rm then}& \left
   [\begin{array}{llllll}s_m=\s+d &\II \ \  \s+1\notin S\ {\rm and} \ c'=d      \\
   s_m\leq \s+d-1&  {\it otherwise}. \qquad(\ref{interv1})  \end{array}\right.\\
   \\
   
 {\rm If}&\left\{\begin{array}{llll} \s\leq  2d  '-d  \\\
  \left [\begin{array}{l}\s +2, d'\end{array}\right] \cap\nat\subseteq S\\
  
  \end{array}\right. & {\rm then}&  s_m= \s+d.\qquad(\ref{interv1})  
 \end{array}$$
 Further, if $H$   denotes the subset of gaps of $S$ inside the interval $[c-e,c'-1]$    and   $\tau$ is the Cohen-Macaulay type of $S$,   \  the exact value or good estimations for $s_m$  are given in the following cases. 
$$\begin{array}{llllcc}
 {\rm If} & H=\emptyset,& {\rm then} & s_m=\s+d \quad (\ref{H01})\\
  {\rm If} & H\ is \ a\ non \ empty\ interval,& {\rm then} & s_m=\left[\begin{array}{llll} 2d'& {\it if}  \ \ \s\geq 2d'+1-d \hspace{1.2cm} \\
  \s+d& {\it otherwise} \quad (\ref{H01}) \end{array}\right.\\
  {\rm If} &S\ is\ associated\ to \ a \  \  Suzuki \ curve, & {\rm then} &  s_m=\s+d\quad(\ref{suzu}).\\
  \\   
  {\rm If} &\n H\leq 2,\quad {\rm see}\quad (\ref{H2}).\\
  \\
  {\rm If} & \e\leq 3,\qquad {\rm see}\quad  (\ref{casol2}),(\ref{3}).\\
  \\
  {\rm If}&\tau\leq 7,\qquad {\rm see}\quad(\ref{tau}).\\
  \\
  {\rm If}&e\leq 8,\qquad {\rm see}\quad(\ref{eminore8}). \end{array}$$
  $$\begin{array}{llllc} {\rm If}&S\ is \ generated\ by\ an\ Almost\ Arithmetic\ Sequence\ and\   embdim(S)\leq 5, \  {\rm see} \ (\ref{AAS}).  \ \quad\end{array}$$

  \section{Semigroups:    invariants  and   relations.} 
  We begin by giving the setting of the paper. 
\begin{set} \label{set1}  
{\rm In all the article we shall use the following notation. Let $\nat$ denote the set of all nonnegative integers. A {\it numerical semigroup} is a subset $S$ of
$\nat$ containing 0, closed under summation and with finite complement in $\nat$; we shall always assume $S\neq \nat$. We denote the elements of $S$ by  $\{s_i\}, \  i\in\nat $, with  
 \ \ $  s_0=0  < s_1     <.  .  .< s_i  <s_{i+1}...  $. 
 \\    We set \     $S(1):=\{b\in\nat\ |\ \
b+(S\setminus\{0\})\subseteq S\}$}
\end{set}

     \noindent We list below the   invariants related to a  semigroup $S\subset \nat$  we shall need in the
sequel. 
\vspace{-0.1cm}$$\begin{array} {cll}
 e&:=& s_1>1,  {   \rm   \   the \ }   multiplicity\ {\rm of} \ S.\\ 

  c& :=& min \ \{   r\in S\ |\  r+\nat\subseteq S \}, {  \rm   \  the \ }   conductor {\rm \ of  \ }S  \\
  
 d& :=&    {\rm the\ greatest\ element\ in\ {\it S} \ preceding\ {\it c},  }\   the \ dominant\ {\rm of}\ S\\ 

c  '&:=& max\{s_i\in S \ |\ s_i\leq d\ \  and\ \ s_i-1\notin S\},\   {\rm   the}\   subconductor {\rm \  of\ } S  \\ 

  d  ' &  :=&   {\rm the\ greatest\ element \ in \ S\ preceding\ } c  ', {\rm
\    when }\ \ c'>0 \\
k&:=& d-c  '\\ 
q&:=& d-d  '\\

 \ell& :=& c-1-d, {\rm \   the\ number\ of\   gaps\ of\ } S\ {\rm greater\ than} \ d\\  
g  &:=& \n(\nat\setminus S) , \ {\rm the} \ genus {\rm \ of\  } S\ (={\rm the\ number \ of\   gaps\ of }\ S) \\
 \tau&:=& \# (S(1)\setminus S),  { \ \rm      the \ }   Cohen {\rm -} Macaulay \ type   {\rm\ of}\ S  \\
 
\end{array}\vspace{-0.1cm}$$

 \noindent  Since $c-e-1\notin S$ we have $  c-e\leq c' $;   we define the following sets
 \vspace{-0.1cm}$$\begin{array} {cll}
\

    H\ &:=& [c-e,c']\cap\nat\setminus S\ \ \subseteq \nat\setminus S\hspace{5.7cm}   \\
   S  '&:=& \{s\in S\ |\ s\leq d  '\} \subseteq S.\\
 \end{array}\vspace{-0.1cm}$$
 \noindent Further  \     we shall describe any semigroup $S$ \ with $c'>0$ \  as follows:
\vspace{-0.3cm}$$\begin{array}{ll}\hspace{7  cm}\ell \ {\rm \scriptstyle{  gaps}}  
 \vspace{-0.1cm}\\
  S=\{ 0,   *\ \!.\! \ .\!\ .\ *,\ e,   \  .\!\ .\!\ .\!\   ,\  d  ', *\ .\!\ .\!\ .\!\ * ,\ c  '\ \longleftrightarrow\ d,  *\ .\!\ .\!\ .\!\ *,\
c\rightarrow  \}=   S  '   \cup   \{  \ c  '\ \longleftrightarrow\ d,  *\ .\!\ .\!\ .\!\ *,\
c\rightarrow  \},\end{array}$$
where  \     
  $\lq\lq * "$ \  indicates    {\it gaps}, \ $ \lq\lq *\ .\!\ .\!\ .\!\ *"$ {\it interval of all gaps}, \  and \ $\lq\lq \longleftrightarrow
"$ \ {\it intervals without     gaps}.\\

\noindent Moreover     for $s_i\in S$   we  fix the following notation. 
  $$\begin{array}{lcllll}N(s_i)  &:=&   \{(s_j,s_k)\in S^2\ |\  s_i=s_j+s_k\}; &\nu(s_i)\!&:=&   \n N(s_i),\\
  &&&\eta(s_i)\!&:=&   \nu (s_{i+i})-\nu (s_i).\ \\
    d_{ORD}(i)\!&:=&min \{\nu(s_j)\ |\ j>i\},  \ {\rm the \ }  order\ bound .
\vspace{0.2cm}\\
  A\ \! (s_i)&:=&\{(x,y), (y,x)   \in N(s_i) \ |\    x<c  ',\     c  '\leq   y\leq d \}; 
  
& \alpha(s_i)&:=&\n A(s_{i+1})-\n A(s_i). \vspace{0.2cm}\\

B\ \!(s_i)&:=&\{(x,y)\in N(s_i)\ |(x,y)\in[c  ',d]^2\ \} ;& \beta(s_i)&:=&\n B(s_{i+1})-\n B(s_i). \vspace{0.2cm}\\

   C\ \!(s_i) \!&:=&   \{(x,y)\in S  '^2\cap N(s_i)\};
& \gamma(s_i)\!&:=&   \n C(s_{i+1})- \n C(s_i).\vspace{0.2cm}\\
           D\ \!(s_i)\!  &:=&    \{ (x,y),\ (y,x)\in N(s_i)\ |\   \  x\geq c \};    &\delta(s_i)&:=&\n D(s_{i+1})-\n D(s_i).

  \end{array}$$
  Now we recall some definition and former results     for completeness. 
First,   a  semigroup $\ S\ $ is called    
 $$\begin{array}{llll}   {\it ordinary}   &if&   S=\{ 0\}\cup \{ n\in \nat, \ n\geq c\}   \vspace{0.2cm}\\

 {\it acute}&if&  either\  S\ is\ ordinary, \ or    \  \  c,d,c  ',d  ' \ \ satisfy  \  \    \ c-d\leq
c  '-d  '  \quad  {\rm \cite[Def. \  5.6]{ba}}.\end{array}$$

 \begin{defin} \label{defmt}   We define the invariants $\ \s$,  m   and   t \ as follows.\enu
 \item[] 
 $    \s\ \!\ :=\ max\ \{s\in S $   such
that \         $s-\ell \notin S\}$.
\vspace{0.1cm} \\   
$   t\ \!\ :=\ d-\s$.
\vspace{0.1cm} \\   
$m:=\ min \ \{j\in\nat $ such that the sequence $     \{\nu(s_i)\}_{i\in\natsmall}  $ is non-decreasing for $i>j\}\vspace{0.1cm} $   \\
$(  m>0\II \nu(s_m)>\nu (s_{m+1})$ and $\nu ( s_{m+k}) \leq \nu (s_{m+k+1}) , \ $   for each   $k\geq 1).$
\denu
 \end{defin}

 \begin{thm}  \label{niedi}    Let $S=\{s_i\}_{\ i\in \nat}$ be as in Setting \ref{set1}.  
  \begin{enumerate}[{\rm(1)}] 
\item  $ \nu(s_i)= i+1-g$,\ \  \ for every \   $s_i\geq 2c-1$. \ \ {\rm   \cite[Th. 3.8]{kp} }
 \item   $ \nu(s_{i+1})\geq \nu(s_i)$, \ \ for every   $   s_i\geq 2d+1$. \ \ {\rm   \cite[Prop. 3.9.1]{ot1} } 
\item If $S$ is an ordinary semigroup, then $ \ m=0$. \ \ {\rm  \cite[Th. 7.3]{ba} }
\item If $\ \s \geq d'$, then \ $s_m=\s+d$ \ {\rm \cite[Th. 4.1,\ Th.4.2]{ot2}}.
\item[] In particular: 
\enua 
\item[$(a)$] if $\ t\leq 2$, then $\ s_m=\s+d $,
\item[$(b)$] if $S$ is an acute semigroup, then \ $s_m=\s+d$, \  with 
\begin{enumerate} \item   either \ $d-c  '\geq \e-1$,  $ \ s_m= 
c+c  '-2  = \s+d ,$
\ 
\
\item   or \qquad   $\s=d\ \ ( s_m=2d)$.  
\ {\rm  \cite[Prop. 3.4]{ot1} }.
\denu
\denu
\item If \ $c  '\in\{c-e,c-e+1\}$, then $S$ is acute. {\rm \cite[Lemma
5.1]{ot2}}. 
\end{enumerate}
\end{thm} 

\begin{rem}\label{remacuti} {\rm (1) By the definition of  \ $\s\ $ it is clear that:\\ \centerline{
 $ s-\e\in S \ \ for \ each \ \ s\in S\ \ such\   that \ \ \s<s\leq d.\quad$}
(2).  Theorem \ref{niedi} implies that $\ 0<s_m\leq 2d\ $  
for every non-ordinary semigroup.\\  
    (3) The condition $(a)$ of (\ref{niedi}.4) does not imply $S$ acute: see (\ref{minor2d}.2). Analogously there exist non-acute semigroups satisfying the conditions $(4.b, i-ii)$,  as shown in  Example   \ref{ex1}.2. }\end{rem}

  We complete this section with some general relation   among the invariants defined above.
  \begin{prop}   \label{minor2d}  {\rm \cite[Prop. 2.5]{ot2}}
 Let $c  '=c-e+q,\ $ with $q\geq 0$. Then 
\enu
    
   \item $e\leq 2\e+t+q$.
   \item The following conditions 
\enua 
\item $ d-c  '\geq \e-1  \ \ \  ($ i.e. $ \ c+c  '-2\leq 2d)$.

\item $\s-\ell =c  '-1$.
\item $c+c  '-2=\s+d$.
\item $e=2\e+t+q$

\item[]  are equivalent and imply

 \enu    
 \item[$(i)$]  $c  '\leq \s\leq d\qquad(\I s_m=\s+d)$.
 \item[$(ii)$] S is acute $\II d-d  '\geq 2\e+t$.
 \denu
 \denu
\denu
     \end{prop}
 Proof. (1)  By
(\ref{defmt}.1) we have
$\s-\e\leq c  '-1=c-e+q-1$, then
$\s-\e\leq d+\e-e+q$ and so $e\leq 2\e+t+q$. \\
(2) The equivalences  $(2.a ) \II (2.b )\II (2.c )$ are proved in \cite[Prop.
2.5]{ot2}.   Clearly the equality
$e=2\e+t+q$ holds if and only if $\ \s-\e=d-t-\e=c  '-1$. Further:\\
  ($i) $  is    obvious by (2.$b$).\\
($ii) $ 
 If (2.$b) $   holds, then
$d-d  '-(2\e+t)=\s-\e-d  '-\e=(c  '-d  ')-(\e+1)=(c  '-d  ')-(c-d).$ Then $S$ is acute
$\II d-d  '\geq 2\e+t$.

  \begin{prop} \label{rec0}  The following facts hold.  
\enu \item \enua \item  If $\ 0\leq h<e$ and $d-h \in S$,    then \ $e\geq
h+\e+1$.
\item  If $\ s,s  '\in S, \ \ s\geq c-e,\ \ s-\e\leq s  '<s $, then
$\ \ s  '\geq c-e$.
\denu 
\item   $ \ \s\geq c-e$ \ \  $($equivalently, $\ \ e\geq t+\e+1 )$.
\item  Let $t>0$ and let  $\ s':=min\{s\in
S\ |\ s>\s\}$. Then 
\item[] $e\geq 2\e+1+d-s'\geq 2\e+1
\ \ ($equivalently, $s'\geq c-e+\e )$.
\item[] In particular, 
$\s+1\in S\ \I   \ e\geq 2\e+t$.
  \item  One of the following
conditions hold
\item[]  $ \begin{array}{llll}
 (a) & \s-\e> c-e-1 \ \ (equivalently\  \ e>2\ell+t,\ equivalently  \ \ \s-\e\in H)\\
 (b) & \s-\e= c-e-1\ \ (equivalently\   \ e=2\ell+t)\\

(c)&c-e-\e\leq   \s-\ell< c-e -1 \ \ (equavalently\   \   e<2\ell+t) \ 

\end{array}  $
\item    Assume \ $e<2\ell+t$, then $:$
\enua \item  either $\ \s\leq d   '$, \  or   $\ t=0 $.
\item in case
$\ \s\leq d  '\ $ we have: 
 $  [\s+1, c-e+\ell -1]\cap S=\emptyset,$ \ \  
       $\   \n H \ \!\geq 2\ell+t-e>0$.
\denu
 
\denu
\end{prop}
Proof. (1.$a$) \  We have $d<d-h+e \in S$. Hence $d-h+e\geq c=d+\e+1$.\\
(1.$b$) \  If $s\geq c$ we get $s  '\geq c$. If $s\leq d,$  let  $s  '=d-k,\
s=d-h\geq c-e$ (hence $h\leq e-\ell-1) ,$ then  $ d-\e-h\leq d-k \I  k\leq h+\e
\leq e-1 $. Now apply $(a)$.\\
(2) \  Let $d=c-e+h\e+r,$ with $h\geq 0,\ 0\leq r<\e$ (recall that  
$c-e\leq d
\
\!$). If $\s<c-e$, first note that by (\ref{remacuti}.1)  we get   
$d-h\e\in S,\ d-(h+1)\e\in S$; further we get $ c-e-\ell \leq
d-(h+1)\ell <c-e$, a contradiction 
  because $[c-\ell-e, c-e-1]\cap S=\emptyset$ for every semigroup.
 \\ (3) \  By (\ref{remacuti}.1), $ \s<s'\leq d\I  s'-\e\in S$ and  so $
s'-\e+e\in S$. Since $ c-e\leq \s<s'$, \  we get
$s'-\e+e> c- \e=d+1$; it follows that \ 
$s'-\e+e\geq c$.\\ 
  (4) \ Since $c-e-1\notin
S$, the  statements are  almost immediate by (2).   \\
(5) \  In
case   $e<2\e+t$, by (3) we have $\s+1\notin S$,
therefore we have $\s\notin [c  ',d)$, hence $(5.a)$ holds. \\
$(5.b)$  Since $\s\geq c-e$ (2), we get    $[ \s-\e+1, c-e  -1]\cap
\nat\subseteq[c-\e- e +1, c-e  -1]\cap \nat\subseteq
\nat\setminus S$. We deduce  
$[\s+1, c-e+\ell -1]\cap
\nat\subseteq H$   by (\ref{remacuti}.1).   
\quad$\diamond$ 
   \begin{coro}\label{cororecall}  Assume $\ \s<d \ \ ($i.e. $t>0)$. Then 
   \enu\item  If \  $\s\leq d'$, then \  $d-c'\leq \e-2$, \   $d-d'\leq \e$, \  \ $c'\geq c-e+2$.

\item Let $c  '=c-e+q,\ $  with
$q\in \{0,1\};$ then  \ 
  $d-c  '\geq \e-1\ $ and $\  e=2\e+t+q$.
\item If $\ d-c  '\leq \e-2,$ \ then 
\enua \item $ d-c  '+2\leq \  d-d  '\leq \e\leq e-3-(d-c  ')$.
\item If \ $\s\geq 2d  '-d$, \ then $\ t\leq 2\e$.

 \denu
 \item If $\s<d  '$ and $e\leq 2\e+t$, then  $\ \n H \leq \e+t-2(d-c  ')-4$.

\denu
\end{coro} 
Proof. (1) By (\ref{minor2d}.2) we see that $\s\leq d'\I d-c'\leq \e-2$  and also $d-d'\leq \e$.
 Further $c  '\geq c-e+2$ \  because  \  $c-e\leq \s $ (\ref{rec0}.2) and \ $\s\leq d  '\leq c  '-2$.\\
 (2) \ By the assumptions and by (5), (4) of Theorem \ref{niedi}, we
have $d-c  '\geq \e-1$.  Then the other statement   follows  by
(\ref{minor2d}.2). \\
  (3.$a$) \  Since $d  '\leq c  '-2$   the first inequality holds for any
semigroup.  We have $d-c  '\leq \e-2$, by   assumption, 
  and   $d-\e\in S$, by (\ref{remacuti}.1). Hence
$d  '\geq d-\e$.  For the last inequality see \cite[Prop. 5.2]{ot2}]. \\
 $(3.b)$ follows by
$(3.a)$ because the assumption means $t\leq 2d-2d  '$.\\
(4)  By
(\ref{minor2d}.2) we have
$d-c  '\leq \e-2$, then by )\ref{rec0}.1-2) and (3$a$),\ \ we deduce that
$\{c  '-\e,...d-\e,\s,d  '\}\subset S\cap[c-e,d  ']$. Hence $\n H\leq
c  '-(c-e)-2-(d-c  '+1)= 2c  '-2d-\e-1-3+e\leq
\ 2c  '-2d-\e-1-3+2\e+t=\e+t-2(d-c  ')-4$.
   \quad$\diamond$ 

\begin{coro} \label{coroH}\enu
\item If \ $c  '\leq \s< d$, then $\ e\geq 2\ell+t$.
\item  If \ $\s\leq d  '$, \ we have:
\enua
 
 \item If \ $e\leq 2\e+t $, then \ $d+2\e-e\in S\II e= 2\e+t$.
\item If \ $H\subseteq[d  '-t+1,c  '-1]$, then \
 $e\leq  2\ell+t $.
 
\item If \ $H=[d  '+1,c  '-1]\cap\nat$ \  and \   $e<2\ell+t$, then 
 $\ \s=d  '$.
\denu 
\denu

\end{coro}
Proof. (1) is immediate by $(\ref{rec0}.3)$ because   $\s+1\in S$.\\
(2.$a$)  
\  Clearly $e=2\ell+t\I  d+2\ell-e=\s\in S$. The converse follows by the assumption and by(\ref{rec0}.5$b$):   \ $c-e+\e-1=d+2\ell-e\in S\I e\geq 2\e+t$; then  $e=2\e+t$.\\
(2.$b$) \  $\s\leq d  '\I d-\ell\in S\I d-\ell\leq
d  '$  by (\ref{cororecall}.1).
\ Hence
\ $\s-\e\leq d  '-t:$ \ \
 now the assumption on $H$ implies $\s-\e\notin
H$, \  i.e., \ $e\leq 2\ell+t$ (\ref{rec0}4).\\
(2.$c$)  \  If $e<2\ell+t$,
we have $\s+1\notin S$ (\ref{rec0}.3); since $c-e<\s+1$ we get
$\s+1\in H$, and so $\s=d  '$.\quad $\diamond$
 
 \begin{ex} \label{ex1} 
  {\rm 
  \enu
\item If $t>0$, for each  $s_i $ such that $\s<s_i\leq d$, we have that $s_i-\ell\in S$, hence $s_i-s_{i-1}\leq \ell$, but it is not true that for each $s_i\in S $ such that $ c-e\leq s_i< d$, we
have
$s_{i+1}-s_i\leq \e$: for instance let $S=\{0, 5\e _e,\! 7\e
_{\s=d-\ell=d  '},\ \! 8\e _d, \ \! 9\e+1 _c\rightarrow\}$.
\item When
$t=0$ the inequality  $e\geq 2\e+1$ (proved in (\ref{rec0}.3) for $t>0$) in general is not true, even for acute
semigroups:     \par
   $S_1=\{0,\ 10_{e=d  '},\  17_{c  '}, 18, \ 19,\ 20_{d } ,\ 
27_c\rightarrow\}:$ \par $\qquad\e=6,\   t=0 \ \  $, $S_1$ is acute with
$d-c  '\leq \e-2,\ e< 2\e$.\par
  $S_2=\{0 ,\ 8_{e},\ 12_{d  '},\ 14_{c  '}, \ 15,\
16_d,\ 20 _c\rightarrow\}:$ \par $\qquad\e=3,\   t=0 \ \   $, $S_2$ is
non-acute  with $d-c  '=\e-1,\ \ e>2\e$. \par
  $S_3=\{0 ,\  7_{e},\  12_{d  '},\ 14_{c  '=d},\  19_c\rightarrow\}:$ \par
$\qquad\e=4,\   t=0, \ \   $ $S_3$ is non-acute with $d-c  '\leq \e-2,\ \
e<2\e$.\par
$S_4=\{0,\ 10_{e=d  '},\  14_d,\ 20_c\rightarrow  \}$\par \qquad $\e=5,
\ t=0,$ $S_4$ is non-acute with $d-c  '\leq \e-2,\ e=2\e.$ 
 \item  When $\s\leq d  '$ we can have every case   $(a),(b),(c)$ of
(\ref{rec0}.4): 
\par $S_5=\{0 ,\ 13_e,\ 15 _{d  '} ,20 d,\ 26_c\rightarrow\}:\e=  t=5
\ \ e< 2\e+t=15 $;\par
\par $S_6=\{0,\ 15_e ,\ 19_{d  '=\s} ,  24 d,\ 30_c\rightarrow\}:\e=  t=5
\
\ e= 2\e+t=15 $;\par
\par $S_7=\{0 ,  26_e,\  28, 31_{d  '}, 33_d,\ 39_c\rightarrow\}:\e=  t=5
\ \ e> 2\e+t=15 $.

 \denu }

 \end{ex}

\section {General results on s$_m$.}
 We saw in \cite{ot2}, that  $s_m=\s+d$,   when $\s\geq d  '$. To give estimations of $s_m$ in the remaining cases we shall   use the same tools as in     \cite{ot2}: we  recall them     for the
convenience of the reader and we add some improvement,  as the general inequalities (\ref{cof}.3) on the difference $\nu(s+1)-\nu(s)$.   Therefore great part of the following  
(\ref{cof}),(\ref{corof}),(\ref{all}) is already proved in \cite[4.1, 
4.2, 4.3]{ot2}. 

\begin{prop} \label{cof} Let     $S  '=\{ s\in S,\
|\ s\leq d  '\}.$ For $s_i\in S$, let $ \ \eta(s_i),\ \alpha(s_i),\ \beta(s_i),\ \gamma(s_i),\ \delta(s_i)$
\ be as in {\rm(\ref{set1})}. 
Then:
\enu 
\item If   $\ \s<d  '$, we have:\\ $s_{i+1}=s_i+1$ \ for $\ s_i\geq \s+d  '-\ell$, in
particular for $s_i\geq 2d  '$.
\item
 
  For each $s_i\in S:$ \   
 $ \ \eta(s_i)=\alpha(s_i)+ \beta(s_i)+ \gamma(s_i)+ \delta(s_i)$. Further
  
\item[] $    \alpha (s_{i})=
\left[\begin{array}{rllll} -2& if \ \ \ \ ( s_{i +1}-c  '
\notin S  '\ and \ \  s_{i}-d\in S  ')\\
   0& if  \ \ \ \  (s_{i +1}-c  ' \in S  '\II  s_{i}-d\in S  ')\\
 2& if  \ \ \ \  (s_{i+1} -c  ' \in S  '           \ and \ \  s_{i}-d\notin S  ').
\end{array}\right.$ 
\item[] $\beta(s_i)=\left[
\begin{array}{rlcc}
0& if\ \ \ \ s_i\leq 2c  '-2  \ \ or\ \   s_{i}>2d\\ 
1& if\ \ \ \ 2c  '-1 \leq s_{i}\leq c  '+d-1 \\
-1& if\ \ \ \ c  '+d \leq s_{i}\leq 2d.\end{array}\right.$  
\item[] $\gamma(s_i)=\left[
\begin{array}{rlcc}
0&\ \  if\ \ \ \ s_i\geq 2d  '+1\\ 
-1&\ \  if\ \ \ \ s_{i} =2d  ' \\
-1 &\ \  if\ \ \ \   s_{i}< 2d  '$ \ and \     \
$[s_i-d  ',d  ']\cap\nat\subseteq S.\end{array}\right.$ 
\item[] $\delta(s_i)=\left[
  \begin{array}{lll}0 &\ \ if &\ \  s_{i+1}-c\notin S,\ \ s_i\leq 2c-1\\
 2&\ \ if&\ \ s_{i+1}-c\in S, \ s_i\leq 2c-1\\ 
1 &\ \ if&  \ \ s_i\geq 2c\ .\\  
 \end{array}\right.$

 \item   Let  $\   s=2d   -k <2d  $ and $ s+1\in S$, \ then$:$ 
  \enua
\item  $-\displaystyle{\Big[\frac{k}{2}\Big]}-1\leq \nu(s +1)-\nu(s)\leq  \displaystyle{\Big[\frac{k+5}{2}\Big] }$. 
\item 
If  $ \ s=2d'  -h <2d'   $, then  \  \  $-\displaystyle{\Big[\frac{h}{2}\Big]}-1\leq \gamma(s)\leq  \displaystyle{\Big[\frac{h+1}{2}\Big]}$. 
\denu
\denu
\end{prop}
Proof. (1) \ By  assumption and by (\ref{rec0}.2)  we have $c-e\leq
\s<d  '$ and so 
$  d  '-\ell\in S$. It follows $  d  '-\ell\geq e$ because $d  '\geq e\geq
\ell+t+1$ (\ref{rec0}.2). Hence
$s\geq \s+d  '-\ell\I s\geq c$.\\ 
 (2) \ By \cite[(3.3)...(3.7)]{ot2}
we have only to prove the last two statements for $ \gamma$. Let
$s=2d  '-h\in S,\ h\in\nat, \  s+1=2d  '-h+1$ and assume
$[d  '-h,d  ']=[s_i-d  ',d  ']\cap\nat\subseteq S$. Then: 
$C(s_i)=$\\ $\{(d  '-h,d  '),(d  '-h+1,d  '-1),(d  '-h+2,d  '-2),...,(d  '-1,d  '-h+1),(d  ',d  '-h) \}$\\
$C(s_{i+1})=\{(d  '-h+1,d  '),(  d  '-h+2,d  '-1),..,(d  '-1,  d  '-h+2),(d  ',  d  '-h+1)\}$\\ 
it follows that $\gamma(s_i)=\n  C(s_{i+1})-\n C(s_i)=h-(h+1)=-1 $.\\
(3.$a$) To prove the inequalities for $s=2d -k$, divide the interval $[d -[k/2],d ]\cap\nat$ in
 subsets\\
 \centerline{ $\Lambda_j:=H_j\cup S_j ,\ \  \ j=1,...,j(s)$, with $H_j\subseteq \nat\setminus S, \ S_j\subseteq S,\ $ \ $S_j=[a_j,b_j ]\cap\nat$  \ interval} such that $b_j+1\notin S$ and $\ H_j\neq \emptyset $, \  if $j > 1$ \ (i.e. $a_{j-1}\notin S$ for $j>1$, $H_1=\emptyset \II  a_1= d -[k/2]\in S)$. Hence \\
		 \centerline{$\Lambda_j= [***\ a_j \longleftrightarrow b_j ].$} \  
	Let $N(s )_j:=N(s )\cap\{(x,y),(y,x) \ |\ y\in S_j\}   $: we have 
	$N(s )={\bigcup  }_j N(s )_j \cup D(s )$. Hence:\\ \centerline{ $\nu(s+1)-\nu(s )=(\sum_j n_j )+\delta(s),$ \ where  $n_j=\n  N(s +1)_j-\n N(s )_j $.} \\Further:   \ $-2\leq n_j\leq 2$.\  This fact follows by   the same argument used to prove the  formulas for $\alpha(s_i),\ \beta(s_i)  $ recalled in statement (2) above.  Since \  $0\leq \delta(s )\leq 2 $ (see (2) above) \ we conclude that \\ \centerline{ $(*)\qquad\quad\ \ -2j(s)\leq \nu(s+1)-\nu(s )\leq 2j(s) +2$.\qquad\qquad\ \ }
 More precisely, to evaluate the largest and lowest possible values of $\nu(s+1)-\nu(s )$, with $s=2d-k$, we consider separately four cases:  $\left[ \begin{array}{lll}
 (A)& k=4p\\
 (B)& k=4p+1\\
 (C)& k=4p+2\\
 (D)& k=4p+3.\\
 \end{array}\right.$\\
 In each case we can see that  \ $j(s)\leq p+1= \displaystyle{\Big[\frac{k}{4}\Big]}+1.$ 
 First note that $d\in S$, hence $j(s)$ is maximal when $\n H_j=\n S_j=1$, i.e. \ $[d -\displaystyle{\left[\frac{k}{2}\right]},d ]= [...*\times*\times...*d]$ (where $\times$ means element$\in S$). \\
In each of the above cases we shall find integers $x_1,x_2,y_1,y_2$ such that    $\left\{\begin{array}{cll}
  x_1\leq \nu(s+1)\leq x_2\\
  y_1\leq \nu(s)\leq y_2\end{array}\right.$, then the statement will follow by   the  obvious  inequality $  x_1-y_2\leq \nu(s+1)-\nu(s)\leq x_2-y_1$.
 
 - If either $\ k=4p$, \ or $\ k=4p+1$, then  $j(s)$ is maximal if and only if \\ \centerline{$[d -\displaystyle{\left[\frac{k}{2}\right]},d ]= [d-2p*...*d-2 *d]$, \ with $j(s)=p+1$.}
 Note that when $j(s)=p+1,$ \ then $  \ 1\leq \n \big ( N(s)\setminus D(s)\big)\leq 2p+1$ because $(d-2p,d-2p)\in N(s)$; further we have $p\leq j(s+1)\leq p+1 $ and so $\ \ 0\leq \n N(s+1)\leq 2p+4$.\\
 If $\ k=4p$,  \  we have $1\leq \n\big ( N(s )\setminus D(s )\big)\leq 2p+1,$ since $(d-2p,d-2p)\in N(s)$, further $j(s+1)=p$, hence $0\leq \n\big ( N(s+1)\setminus D(s+1)\big)\leq 2p$. \ \   \vspace{0.2cm}\\
 \centerline
 {$-\displaystyle{\left[\frac{k}{2}\right]}-1=-2p-1\leq \nu(s+1)-\nu(s)\leq 2p +2-1<  \displaystyle{\left[\frac{k+5}{2}\right]}$.}\vspace{0.2cm} 
   If $\ k=4p+1$, we have \  $0\leq \n\big ( N(s )\setminus D(s )\big)\leq 2p+2,$  further\ \ $s+1=2d-4p$,  \ therefore \ $1\leq \n\big ( N(s+1)\setminus D(s+1)\big)\leq 2p+1 $. We obtain: \vspace{0.2cm}\\
 \centerline
 {$-\displaystyle{\left[\frac{k}{2}\right]}-1=-2p-1\leq \nu(s+1)-\nu(s)\leq 2p +3= \displaystyle{\left[\frac{k+5}{2}\right]}$.}

 - If either $\ k=4p+2$, or $k=4p+3$, then $ \displaystyle{\left[\frac{k}{2}\right]}=2p+1$,\\ analogously we get  $j(s)=p+1$ maximal if   
$$ \left[ d -\displaystyle{\left[\frac{k}{2}\right]},d\right] =\left[\begin{array}{lll} 
     \left[\right.*d-2p*...*d-2 *d\left]\right. \ \ \  or \\
         \left[\right. d-2p-1   ...  *\times \times*... d\left]\right.  ({\rm \ with\ one\ and\ only\ one} \ {j_0}\  {\rm such\ that\ }  \n S_{j_0}=2). \end{array}\right.$$
 If   $\ k=4p+2$, in the first subcase  we get  $     0\leq \big ( N(s)\setminus D(s)\big)\leq 2p+2$,\ \ and \ $ 0  \leq \n\big ( N(s+1)\setminus D(s+1)\big)\leq 2p$ because $(d-2p-1,d-2p)\notin N(s+1)$.
 Hence \vspace{0.2cm}\\  \centerline
 {$-\displaystyle{\left[\frac{k}{2}\right]}-1=-2p-2\leq \nu(s+1)-\nu(s)\leq 2p +2<   \displaystyle{\left[\frac{k+5}{2}\right]}$.} 
  \vspace{0.2cm}
 In the second subcase  we get  $    1\leq \n \big ( N(s)\setminus D(s)\big)\leq 2p+2$,\ \ because $(d-2p-1,d-2p-1)\in N(s )$ and $ 0  \leq \big ( N(s+1)\setminus D(s+1)\big)\leq 2p$ since $(d-2p-1,d-2p)\notin N(s+1)$.
 Hence\vspace{0.2cm}\\  \centerline
 {$-\displaystyle{\left[\frac{k}{2}\right]}-1=-2p-2\leq \nu(s+1)-\nu(s)\leq 2p +1<   \displaystyle{\left[\frac{k+5}{2}\right]}$.}\vspace{0.2cm} 
If   $\ k=4p+3$, in the first subcase  we get  $     0\leq \big ( N(s)\setminus D(s)\big)\leq 2p+2$,\ \ and \\ $ 0  \leq \n\big ( N(s+1)\setminus D(s+1)\big)\leq 2p+2$ because $(d-2p-1,d-2p)\notin N(s+1)$.
 Hence \vspace{0.2cm}\\ \vspace{0.2cm} \centerline
 {$-\displaystyle{\left[\frac{k}{2}\right]}-1=-2p-2\leq \nu(s+1)-\nu(s)\leq 2p +4=   \displaystyle{\left[\frac{k+5}{2}\right]}$.}  
    In the second subcase  we get  $    0\leq \n\big ( N(s)\setminus D(s)\big)\leq 2p+2$ \ \  and $ 0  \leq \n\big ( N(s+1)\setminus D(s)\big)\leq 2p+1	$ because $(d-2p-1,d-2p-1)\in N(s+1)$.
 Hence \vspace{0.2cm}\\ \vspace{0.2cm} \centerline
 {$-\displaystyle{\left[\frac{k}{2}\right]}-1=-2p-2\leq \nu(s+1)-\nu(s)\leq 2p +3<   \displaystyle{\left[\frac{k+5}{2}\right]}$.} \\
 (3.$b$) The proof is quite similar to the above one: since $\gamma(s)=\n C(s+1)-\n C(s)$,  we do not need to add $\delta(s)$ and so     formula $(*)$  becomes   \\
 \centerline{ $\ \ -2j'(s)\leq \gamma(s)\leq 2j'(s)  $, }
 where $j'(s)$ is the number of subset $\Lambda_j$ as in (3.$a$) contained in the interval $[d' -\displaystyle{\left[\frac{h}{2}\right]},d' ]\cap\nat$.\\ Then it suffices to proceed as above.\quad $\diamond$

 \begin{ex}{\rm The bounds found in (\ref{cof}.3$a$) are both sharp. To see this fact,  consider \\ $S=\{0,10_e,20_{d'},30_d,40_c\rightarrow\}$ an the elements 
 $s= 2d-1=59$, $s+1=2d=60$. By a direct computation we easily get:   \  $\nu(s+1)-\nu(s)=3= \displaystyle{\Big[\frac{k+1}{2}\Big]+2}$ \ (with $k=1$), \ and \\ $\nu(s+2)-\nu(s+1)=
	 -\displaystyle{\Big[\frac{k}{2}\Big]}-1$ \ (with $k=0$).  }
	 \end{ex}
\begin{prop} \label{corof} Let $\0$ mean
$\notin S  '$ and $\times$ mean $\in S  '$ \  $($recall that for $s\leq d  ' $,
  we have $s\in S\II s\in S  ')$.    The
following tables  describe   the difference \ $\eta(s_i)= \nu(s_{i +1}) -\nu(s_{i})$
for 
$s_{i}\in S, \ s_i<2c   $ in function of $\alpha,\beta,\gamma,\delta$. 
\enua 
\item If $s_i<2c:$\\
$\left[\begin{array}{cccrrlcccc}	 
s_{i+1}-\!c	 &	s_{i}-d	&s_{i+1}-\!c  ' & \alpha	&	\ 
	\ 	\delta&\quad\ \ \eta(s_i)\\

\notin S	&	 	\times	&	\0	&	-2	&		 	0	&\ \ \ 	\beta+\gamma-2  	\\
\notin S	& 	\0	&	\0	&	0	&	 	0	&\ \ \ 	\beta+\gamma  	\\
\notin S	&	 	\times	&	\times	&	0	&		 	0	&\ \ \ 	\beta+\gamma	\\
\notin S	&	 	\0	&	\times	&	2	&	 	0	&\ \ \ 	\beta+\gamma+2	\\

\in S	&	 	\times	&	\0	&	-2	&		 	2	&\ \ \ 	\beta+\gamma	\\
\in S	&	 	\0	&	\0	&	0	&		 	2	&\ \ \  \beta+\gamma+	2	\\
\in S&	 	\times	&	\times	&	0	&		 	2	&\ \ \ \beta+	\gamma+2	\\
\in S	&	 	\0	&	\times	&	2	&	 	 	2	&\ \ \ \beta+\gamma+	4	\\
\end{array}\right]$.\\

More precisely we have the following subcases.

\item  If    $s_{i}\leq 2d  '-1 ,$ then \  $\ \beta=0$:\par
 
$\left[\begin{array}{cccrrrcccc}	 	s_{i+1}-\!c	 &	s_{i}-d	&
s_{i+1}-\!c  ' &
\alpha	&	\ \beta	&	\ 	\delta& \quad\eta(s_i)\quad\\
\0	&	 	\times	&	\0	&	-2	&	0	&	 	0	&	\gamma-2  	\\
\0	&	 	\times	&	\times	&	0	&	0	&	 	0	&	\gamma	\\
\0	& 	\0	&	\0	&	0	&	0	& 	0&\gamma  	\\
	\times	&	 	\times	&	\0	&	-2	&	0	&	 	2	&	\gamma	\\
\0	&	 	\0	&	\times	&	2	&	0	 	&	0	&	\gamma+2	\\

	\times&	 	\0	&	\0	&	0	&	0	&	 	2	&\gamma+	2	\\

	\times&	 	\times	&	\times	&	0	&	0	&	 	2	&	\gamma+2	\\
	\times&	 	\0	&	\times	&	2	&	0	&	 	2	&\gamma+	4	\\
\end{array}\right]$.

  \item  If    $s_{i}= 2d  '  , $ then 
\  $\ \beta=0,\ \  \gamma=-1$:\par
 
$\left[\begin{array}{cccrrrcccc}	 	s_{i+1}-\!c	 &	s_{i}-d	&
s_{i+1}-\!c  ' &
\alpha	&	
\ \beta	&	\ 	\delta& \quad\eta(s_i)\quad\\

\0	&	 	\times	&	\0	&	-2	&	0	&	 	0	&	-3  	\\
\0	&	 	\times	&	\times	&	0	&	0	&	 	0	&	-1	\\
\0	& 	\0	&	\0	&	0	&	0	& 	0	&	-1  	\\

\times	&	 	\times	&	\0	&	-2	&	0	&	 	2	&	-1	\\ 
\0	&	 	\0	&	\times	&	2	&	0	 	&	0	&	\ \ 1	\\
\times	&	 	\0	&	\0	&	0	&	0	&	 	2	&\ \ 1	\\
\times&	 	\times	&	\times	&	0	&	0	&	 	2	&	\ \ 1	\\
\times	&	 	\0	&	\times	&	2	&	0	&	 	2	&\ \ 3	\\

\end{array}\right]$.

\item  If    $s_i\in [2d  '+1,   c  '+d-1] , $ then
\  $\ \beta\in\{0,1\},\ \  \gamma=0$: \par $ \nu(s_{i+1}
)<\nu(s_{i})$ if and only if the following row is satisfied\par
 $\left[\begin{array}{cccccrr}	 	s_{i+1}-\!c	 &	s_{i}-d	&
s_{i+1}-\!c  ' \\ 
\0	&	 	\times	&	\0	
\end{array}\right]$. 
\item If $  s_i\in [c  '+d,2d],$  then \ $\beta=-1, \ \gamma=0,\ s_i-d\in
S\setminus S  ', s_{i+1}-c  '\notin S  '$, then\\
  $\nu(s_{i+1})<
\nu(s_i)\II s_i-\ell-d\notin S$.

\denu
\end{prop}
The next theorem collects with some upgrades the results  \cite[Th. 4.1,\ Th.4.2,\ Th. 4.4]{ot2}: statement (1) improves \cite[Th.4.2]{ot2}, the last part of   (5) is new.
\begin{thm} \label{all}With Setting \ref{set1}, the following inequalities hold.
\enu\item[$(0)$] If \ $ \s\geq 2d'-d ,\ \ then \ \ s_m\leq  \s+d ;$ \ \ \ if \ $\s< 2d'-d ,\ \ then \ \ s_m\leq 2d'.$
 \item[] More precisely 

\item   If \ \   $\s\geq d  '+c  '-d ,\ \ then \ \ s_m= \s+d $.
\item   If \ \   $\s= d  '+c  ' -d-1,\ \ then \ \ s_m\leq \s+d-1 $.

\item If \  $\ 2d  '-d < \s< d  '+c  '-d -1$, let
\item[] 
$U:=\{
\sigma\in
  [2d  '+1-d,\ \s]\cap S\ \ |\   \ \sigma-\ell\notin S,\ \sigma+d+1-c  '\notin S  \}$: 
  \enua

\item  if \ $U\neq \emptyset$,  then \ \ $s_m=d+max\ U$, 
\item[] in particular 
  $s_m=\s+d\II \s+d+1-c  '\notin S$,
\item if \ $U=\emptyset,$ \  then $s_m\leq 2d  ' $.
\denu 
 \item If \ $ \s= 2d  '-d $, \ \ then \ \ $s_m=\s+d$.
 \item If \ $ \s<2d  '-d$, then $s_m\leq 2d  '$,  more precisely:
  
\item[]  $s_m = 2d  '  \II 2d  '$ satisfies either row 3 or  row 4 of
Table \ref{corof}  ($c)$.
 
\item[]  In the case  $\s+d+1-c  '\notin S:$
\enua
 \item if $\ 2d  '-d-2\leq \s\leq 2d  '-d-1, $\  then $\ \s+d\leq  s_m\leq 2d  '$
 \item if \ $\s =2d  '-d-j,\ \ j=3,4  $ and $\{d  '-j,...,d  '-1\}\cap S\neq
\{d  '-j+1\}$ \ \ then  \ $\   s_m\geq \s+d $.  

 \denu
\denu 
\end{thm} 
Proof. (0) is proved in \cite[(4.4.1),(4.4.3)]{ot2}.\\
Now recall that $\s\geq c-e $ (\ref{rec0}.2),   hence $\s+d+1\geq c+1\in S$; further in  cases (1)   and   (2)  $\s+d+1-c  '\geq d  '$, hence  (1) and (2) 
follow   by  (0) and by Table \ref{corof} ($d)$. \\
 The cases ($3) $ and    $(4)$   
follow easily   by Tables \ref{corof} ($d)$ and  ($c)$.\\
 (5) \ We have $\ s_m\leq 2d  '$ by (0);  further 
$2d  '$ cannot satisfy the first two rows of \ref{corof} ($c)$  since
$\s<2d  '-d$.\\
  By a direct computation we can see that we always have  $\gamma(2d  '-j)\leq 1,$ for $j\leq 2$, while for $j=3,4$ \ \ $\gamma(2d  '-j)\leq 1\II\{d  '-j,...,d  '-1\}\cap S\neq \{d  '-j+1\}$. Now $(a)$ and $(b)$   follow  because \ $\nu(\s+d)>\nu(\s+d+1)$  by Tables \ref{corof}.($b)$ and $  (c)$.
 \quad
$\diamond$\\

 The following conjecture  gives a  lower bound for $s_m$, it is justified by calculations in very many examples. We are able to prove that it holds in many  cases.
 \begin{conj} \label{co} For every semigroup the inequality   \ $s_m\geq c+d-e$ \ holds.
 \end{conj}
First we note that (\ref{co}) holds in the following general cases:
\begin{prop} \label{cong1}  Assume $\left[\begin{array}{ll} either & s_m\geq \s+d  \\ or &s_m \geq 2d  '\ \ and \ \ \s<d  ' \
\end{array}\right. $ \  then $\ s_m\geq c+d-e$.\\  
 
 \noindent In   particular if either \ $\s+d\geq c  '+d  '$ \ or \ $\s+d=2d  '$, \ then $\ s_m\geq c+d-e$.

 \end{prop}
Proof. The first part follows by (\ref{rec0}.2)  and (\ref{rec0}.3). In fact we have\\
 (1) $\s\geq c-e$;\\
 (2) \ in case $ \ \s<d  '$, \ $s_m\geq 2d'$, we have  \ $d  '\geq c-e+\e $ (\ref{rec0}.1$b$) \  and \ $d-d  '\leq \e$ \ (\ref{cororecall}.1). Hence $s_m\geq 2d  '\geq d  '+c-e+\e \geq c+d-e$.
 Now the   particular cases follow   by (\ref{all}. 1...4).
 \quad$\diamond$\\
 \begin{coro} \label{coroall}   
\enu \item  
 If $s_m> 2d  '$, \  then \ \ $s_m-d\in S$.
 \item  If \  $\s=d  '-1$, then \ $s_m=\s+d\II c  '\neq
d$.

\denu
\end{coro}
Proof.(1) follows by (\ref{all}.1) and by Table \ref{corof}.$(d)$.\\
(2) \ If $\ c  '=d$, then $s_m\neq \s+d$, by (\ref{all}.2).\\ 
If
$c  '\neq d$ and $\s=d  '-1$, we have $\s\geq d  '+c  '-d$ then apply (\ref{all}.1).
\quad$\diamond$
 
\begin{prop} \label{interv1}    Assume $\s\leq d  '-2$ and
$[\s+2,d  ']\cap\nat\subseteq S$. Then \ $s_m \leq \s+d :$ \enu  
 
\item  if \ $2d  '-d<\s< d  '+c  '-d $, \ then \ $s_m  \left[\begin{array}{lll}   =\s+d\II \s+1\notin S\ and\ c  '=d\\
\leq \s+d-1,  \ \ otherwise.
\end{array}\right.$
 
 \item  if $\ \s\leq 2d  ' -d$, \ then $s_m=\s+d$.
\denu
                         
\end{prop}
Proof. In case (1),  by applying   Theorem \ref{all} we see that  
 $s_m\leq \s+d$ ; further   $s_m=\s+d\II \s+d+1-c  '\notin
S$. Since $ \s+1\leq \s+d+1-c  '\leq d  ' $ by the assumptions,    we see that $s_m=\s+d\II c  '=d$
and
$\s+1\notin S$.\\
In case (2), by  Theorem \ref{all} we have $s_m\leq 
2d  '$.\\ Now let
$\ \s+d+1\leq s\leq 2d  '$. \  Then by the assumptions we get
\\
\centerline{    
 $\left\{\begin{array}{ll} \s+2\leq   s+1-c  '\leq s -d  '-1<d  ' \\
s+1\in S \ {\rm \ and\ }\ s+1-c  '\in S  '\\
\{s-d  ',...,d  '\}\subseteq S \ \ ({\rm hence}\   \gamma(s)=-1\     (\ref{corof}.2))\\
s-\e-d\in S \ \ (\rm{by \ \ (\ref{remacuti}.1)}).
\end{array}\right.$}
From Tables   \ref{corof} ($b)-(c)$ we conclude that \  $s_m<s$  \ and   also that \ $s_m=\s+d$; in fact\\ $\s\in S,\ \s+d+1-c  '\in S  ',\  \s-\e\notin S,$ further $\
\s+d-d  '\geq \s+2,$ because $d-d  '\geq 2$, therefore $
\gamma(\s+d)=-1 $ by the assumptions and (\ref{corof}.2).\quad$\diamond$

\begin{rem} {\rm (1) Both   situations of (\ref{interv1}.1)  above can
happen, even for $\e=3$ (see the following (\ref{3})):
\par (A) If $\e=3,\ t=5$, $d  '=d-3,$\ $
 \ c  '=d-1$,  (\ref{3}.case A) we have
$s_m<\s+d$.
\par (B) If $\e=3,\ t=5$, $d  '=d-3,\ d-4\notin S$, \ $
 \ c  '=d$,  (\ref{3}.case B) we have
$s_m=\s+d$.
\enu \item[(2)]Assume \ $2d  ' -d<\s \leq d  '+c  '-d-1$ \ and \
$[d  '-\e+2,d  ']\cap\nat\subseteq S$;  then the set $U$ of (\ref{all}.3) is empty. In
fact for each $s\in S,$ such that $2d  '+1\leq s\leq \s+d$, we have $s+1\in S$, and by (\ref{cororecall}.$3(a)$),
$d  '-\e+2\leq 2d  '+2-d\leq s+1-c  '\leq d  '$, therefore $s+1-c  '\in S  '$.
\item[(3)] If $s_m<2d  '\leq \s+d$, then $\left\{\begin{array}{ll} (a)\ 
\s+d+1-c  '\in S
\\
(b)\ \s+d+1-c  '-\e\in S \\
(c)\ \{2d  '-d-\e,2d  '+1-c  '\}\cap S\neq\emptyset .
\end{array}\right.$\ 
\item[] (3.$a$) holds by (\ref{all}.3); in fact the assumptions imply $U=\emptyset$  because $\s-\e\notin S$. \\ $(3.b)$
is clear by $(3.a)$ and by (\ref{remacuti}.1) , since $\s<\s+d+1-c  '<d$.\\
 $(3.c)$ follows by Table \ref{corof} $(c)$).

\item[(4)] The assumption $s_m>2d  '$ in (\ref{coroall}.1) is necessary: 
  for instance if \\ $S=\{0,20_e,21,26,27_{d  '},32_d, 39_c\rightarrow\}$ we 
have
$s_m=2d  '$, with $2d  '-d\notin S$ \\  (we deduce $s_m=2d  ' $ by Table 
 \ref{corof} $ (c $)).\denu }
\end{rem}
\begin{prop} \label{int0} If \ $\s<d'$ and 
 $[d  '-\e,d  ']\cap\nat\subseteq S$,  
 let $\ h=d-c  ',\ \ q=d-d  '$ be as in {\rm(\ref{set1})} and let
  $\ \sigma:=max\{s\in S, s<\s-\e\}$. Then

\enu 
\item   $ [\s-\e+1,d  ']\cap\nat\subseteq S$  \ and \
 $ \ e\geq 2\e+t $.

\item If $\ 2d  ' -d<\s\leq d  '+c  '-d-1$,  we have
 \enua \item   $q+h+1\leq  t< 2q\ (\leq 2\e), $ 
 \item   For \ $s\in  [\s+d  '-\e+1,2d  ']\cap  S$, we have $\gamma(s)=-1$.
\item If $\ 2d  '-\e-d\in S$, then $\ \sigma\geq 2d  '-\e-d $ and $\
\gamma(\sigma+d)=-1$.
 \item We have  $s_m\leq 2d  '.$

\item Let \ $W:=[\s-2\e+1,2d  '-\e-d]\cap \nat\setminus S$.\\ If \ $W\neq
\emptyset,$ let $h_0:=max\ W$, \ then $s_m\geq h_0+\e+d$.
\item  $s_m< \s+d-\e+1\II  
[\s-2\e+1,2d  '-d-\e]\cap\nat\subseteq S$,
\item[]
 $s_m<\s+d-\e+1\I e\geq 3\e+t. $
\item If \ $[\s-2\e+1,2d  '-d-\e]\cap\nat\subseteq S  ,$ then \ $s_m\geq
\s+d  '-\e+1$.

 \denu 
\denu
  
\end{prop}
Proof.       (1) By the assumptions and by (\ref{remacuti}.1) we have
$[\s-\e+1,d  ']\cap\nat \subseteq S$; the inequality $e\geq 2\e+t$
follows by
 (\ref{rec0}.3)\\ 
(2) \ Statement $ ( a)$ is immediate by the assumption 
$2d  '-d<\s\leq d  '+c  '-d-1$.\\
\indent
   $( b)$ follows by (1) and by (\ref{corof}.2).\\ 
$(c )$ \ By assumption
$2d  '-\e-d\in S
$, further $2d  '-d<  \s$; \ then  $ 2d  '-\e-d< \s-\e$, hence $\sigma\geq
2d  '-\e-d$. Moreover\centerline{ $\s-\e+1\leq d  '-\e=2d  '-\e-d  '\leq
\sigma+d-d  '< (d-\e)+ \s -d  '<2d  '-d  '=d  ';\quad$} then apply  $( b)$ .\\
($d$) \ By (\ref{all}.3) we know that $\ s_m\leq \s+d$. For each
$2d  '<s\leq \s+d$ we have $ d  '-\e<2d  '+1-c  '\leq s+1-c  '\leq \s+d+1-c  '\leq
d  '+c  ' -1+1-c  '= d  '$. Therefore $s+1-c  '\in S$ and $s_m\leq 2d  '$ \  by 
(\ref{all}.3$b)$. \\ 
($e$) and ($f$). \ Note that     $\  s\in
[ \s+d-\e+1,2d  ']\cap S \I \s-\e+1\leq s-d\leq s+1-c  '\leq s-d  '\leq d  '$, hence
$\{s-d,\ s+1-c  '\}\subseteq S  '$ and $\gamma(s)=-1$, by ($b$) and the assumptions.
By Table 
 \ref{corof} ($b$) we get 
\\ \centerline{$\nu(s)>\nu(s+1)\II s+1-c\notin S. $}
Then $(e)$ follows and the equivalence ($f$) becomes  immediate by ($d$), ($e$), recalling that 
$s+1-c=s-\e-d$. 
We get \ $e\geq 3\e+t$ by
(\ref{rec0}.1$-$2), since $d-(2\e+t-1)\in S$ and $2\e+t-1< e$ by
($a$).\\ 
($g$) \ For $s\in [\s+d  '- \e+1,2d  '-\e]\cap\nat  $, we have
$\gamma(s)=-1$ (see ($b$)).   If there exists $\overline{s}\in
[\s+d  '- \e+1,2d  '-\e]\cap\nat, $ \  $\overline{s}  +1-c  '\notin S$, we have  $\overline{s}-d\in S'$ and we have
$s_m\geq  \overline{s}$ by Table  \ref{corof} $(b)$ (in fact $ \overline{s}-d\in S'$ by the assumptions);   the claim
follows.\\ 
Assume on the contrary that $ [\s+d  '- \e+2-c  ',2d  '-\e+1-c  ']\cap\nat\subseteq
S$: then 
\\ \centerline{$[\s -2 \e +1 ,2d  '-\e+1-c  ']\cap\nat\subseteq S.$} In fact  $q+h+1\leq t$
(\ref{int0}.1)
$\I \s+d  '- \e +2-c  '=d-q-\e-t+2+h\leq   d-2q-\e+1= 2d  '-\e+ 1-d$.
We can iterate the algorithm looking for one element
$\overline{\overline{s}}\in [ 2d  '-\e+1 ,2d  '-\e+d+ 1-c  ']\cap\nat$  such
that
$\overline{\overline{s}}+1-c  '\notin S$. If needed we repeat the
argument till we find $s'$ such that $s'+1-c  '\notin S$:
$s'$ surely exists since $\s-\e\notin S$.  \quad
$\diamond$\\

The previous results can be summarized in the following theorem.

\begin{thm} \label{interv} Assume     \  $\ [d  '-\e,d  ']\cap\nat\subseteq
S$. Then     $s_m\geq  c+d-e$.  In particular:
\enu
\item if \ $2d  '-d <\s< d  '+c  '-d $,  we have
 $ \ \ 
c+d-e\leq \s+d  '-\e +1\leq s_m\leq 2d  ' $,  
\item    $\ s_m=\s+d\ $ in the remaining cases.

 \denu
\end{thm}
Proof.  (1) \ Since $\s<d  '$, we have $[\s-\e+1,d  ']\cap\nat \subseteq S$, \ 
$e\geq 2\e+t$ by (\ref{int0}.1). It follows that $\s-\e+1\geq c-e$ because   
 $[c-\e-e, c-e-1]\cap S=\emptyset$ and $\s\geq c-e$ by (\ref{rec0}.2).\\
   The inequalities  follow   by    items $(d),(e),(f),(g)$ of  (\ref{int0}):\\
if the
set $W$ of (\ref{int0}.2$e$) is not empty then we see that  $s_m\geq  \s+d-\e+1\geq 
\s+d  '-\e +3$  by  (\ref{int0}.2$e$), recalling that   $d  '\leq d-2$. \\ If $W=\emptyset $,
  by (\ref{int0}.2$g$) we get $s_m\geq   \s+d  '-\e+ 1$. 
 \\ (2)
follows by (\ref{all}.1) and by (\ref{interv1}.2). In this case $s_m\geq c+d-e$ by (\ref{cong1}).\\ 
  To prove $s_m \geq c+d-e$ in   case (1),  first assume that 
$W=\emptyset$: \ since $d  '\geq d-\e$ (\ref{coroH}.1) and $e\geq 3\e+t$\\
(\ref{int0}.6), we get  
\begin{centerline}{ $ s_m\geq \s-\e+d  ' +1\geq
\s-\e+d-\e +1 =c+d-3\e-t\geq c+d-e. \hspace{3 cm}
$}\end{centerline}
 If $W\neq \emptyset$, since $e\geq 2\e+t$ we get \\ \centerline{$\hspace{0.5
cm} s_m\geq 
  \s+d-\e+1=c+d-2\e-t\geq c+d-e,\hspace{2 cm}$}
\\  further $d>d  '+1\I  \s+d-\e+1> \s+d'-\e+1 $. \quad $\diamond$  \\

 In Case $(g)$ of (\ref{int0}) we can give more precise evaluations  of $s_m$. 

 \begin{coro} \label{int1}   Let \ 
$[\s-2\e +1,2d  '-d-\e]\cup[d  '-\e,d  ']\cap\nat\subseteq S$,    let 
 $\ 2d  '-d <\s< d  '+c  '-d$ \ and let \ $\sigma $ be as in (\ref{int0}) .   
\enu

\item Assume  $\left[\begin{array}{cl}
either&\ \ \ \ \! c  '-\e-t-1\in S 
\\ or &\left\{\begin{array}{ll} 
 c  '-\e-t-1\notin S\\ 
c  '-2\e-t-1\notin S  
\end{array}\right.
\end{array}\right. $,  then \ $s_m\geq \s+c  '-\e -1$ 
\item Assume  $\left\{\begin{array}{ll}
  c  '-\e-t-1\notin S\\ 
c  '-2\e-t-1\in S  
\end{array}\right.$, then$:$  
  
 \enua 
\item  If $\ \sigma \leq c  '-\e-t-2$, \ then \ $  s_m\geq\ \sigma+d$
.

\item  If $\left\{\begin{array}{ll}
\sigma \geq c  '-\e-t \\
\sigma-\e\notin S
\end{array}\right.$, \ then \ $  s_m= \ \sigma+d $. 
  
\item    
If $  \left\{\begin{array}{ll}
 \sigma \geq  c  '-\e-t,   
\ \sigma-\e\in S\ \ 
  and\\
either  \ \ 2d  '- 2\e-d\notin S\\
\ \ \ or\ \    \ \
    2d  '-\e+1-c  '\notin   S
\end{array}\right.$, \ then \ $  s_m\geq 2d  '-\e  .$  
  \denu 
\denu 
\end{coro}
Proof.   
(1)   
Let
$s=\s+c  '-\e-1$. Then $\ \s+d  '-\e +1\leq s\leq d  '+\s-1<2d  '$ and so
$\gamma(s)=-1$ (\ref{int0}.2$b$). Further $ \ s+1-c  '\notin S$ and, by the
assumptions, \  either $\ s-d\in S  '$ \ or \ $(s-d\notin S  '$ and $s+1-c\notin
S  ')$. The claim follows by Table  \ref{corof}  $(b)$.\\
     $(2.a)$ \ From Table    \ref{corof} ($b$) we  get $\nu
(\sigma+d)>\nu(\sigma+d+1)$. In fact we have $\sigma<\sigma+d+1-c  '\leq
\s+c  '-\e -2+1-c  '=\s-\e-1$. Hence
$\sigma+d+1-c  '\notin S$. Further $\gamma(\sigma+d)=-1$ by the
assumption $2d  '-d-\e\in S $ and by (\ref{int0}.2$c$).\\ 
 $(2.b)$ \ In this case for each $s$ such that  $\ \sigma+d\leq s\leq  \s+d-\e$ we have
$s+1-c  '\in S$: in fact 
$\s-\e+1\leq \sigma+d+1-c  '\leq s +1-c  '<2d  '-c  '<d  '$ (\ref{int0}.2$a$).
Moreover we get $\gamma (s)=-1$ by (\ref{int0}.2$b$).  If
$s>\sigma+d$, then
  $s-d\notin S$ hence by Table
\ref{corof}  $(b)$: \  $\nu(s)<\nu(s+1)$ therefore $s_m\leq \sigma+d$ \ and \
$\nu(\sigma+d) >\nu(\sigma+d+1)\II \sigma-\e\notin S$.\\
 $(2.c)$.  Since $d  '>\s $ we have 
 $\gamma (2d  '-\e)=-1 $; in fact 
 $\ 2d  '-\e\geq \s+d  '-\e +1  $ (\ref{int0}.2$b$).
Then the claim follows by using the assumption  $2d  '-\e-d\in S$ and 
Table \ref{corof} ($b$).
\quad$\diamond$\\

 \section {Some particular case.}
In this section  we shall   estimate  or   give exactly the value
of
$s_m$ in some  particular case. Since   $s_m=\s+d$ for each semigroup $S$
satisfying $\s\geq c  '+d  '-d$,  in this section    we shall often assume
$\s<c  '+d  '-d$.
\subsection{Relations between the order bound  and the holes set H.}
 Let $H:=[c-e,c  ']\cap\nat \setminus S$ be as in (\ref{set1}): when    $H$ is an interval we deduce the value of $s_m$,  if  $\n H\leq 2$ and in some other situation   we give a lower bound for $s_m$.
 
\begin{prop} \label{H01} 
 \enu
\item If $H=\emptyset$,  then $c  '=c-e$ and  $S$ is acute with $s_m=\s+d$.
\item Assume that $\s<d  '$. Then the conditions
\enua
\item   $[d  '-\e,d  ']\cap\nat\subseteq S$ \ and \ $e=2\e+t$.
\item $H=[d  '+1,c  '-1]\cap \nat$.
\denu
are equivalent and imply:
$s_m=\left[\begin{array}{lll} 2d  '& if \ \ 2d  '-d+1\leq  \s\leq  d  '+c  '-d-1 \\
\s+d & in\ the\ remaining\ cases.\end{array}\right.$

\denu
\end{prop}
Proof. (1) \  $H=\emptyset\II c  '=c-e$; then apply (\ref{niedi}. 5 and 4).\\
(2 ), ($a)\II(b)$.   ($a$) implies  that \ $[\s-\e+1,d  ']\cap\nat \subseteq S$, \
and $\s-\e=c-e-1 $  (\ref{int0}.1) and (\ref{rec0}.4). Hence (($b$)) holds. On the
contrary,  ($b$) implies \ $[c-e,d  ']\cap\nat\subseteq S$.  Since $c-e\leq d  '-\e$ by
(\ref{rec0}.3), we get  $[d  '-\e,d  ']\cap\nat\subseteq S$, further $e\leq 2\e+t$ by (\ref{coroH}.2$b$). By assumption we also have $\s+1  \in  S$, hence $e\geq 2\e+t$ by (\ref{rec0}.3); then $(a)$ follows.  \par When $2d  '-d
<\s< d  '+c  '-d$,
  by (\ref{rec0}.1) $c-e\leq d'-\e$, and so (\ref{cororecall}.1) we have $ c-e-\e \leq
d  '-\e-(d-d  ') =2d  '-\ell-d< \s-\ell\leq c-e-1$.  We obtain  $s_m\geq 2d  '$ by
(\ref{int0}.2$e$) because     the set
$W$ as in (\ref{int0}.2$e$)  has            
$2d  '-\e-d=max\ W$. Now    $s_m=2d  '$ follows by
(\ref{int0}.2$d$). \ \  For the statement   in the remaining cases see
(\ref{interv}.2).\quad
$\diamond$

\begin{ex} {\rm When $\s>d  '$ the implication $(b)\I(a)$ in (\ref{H01}.2)   is not true in general:
in fact  \ \
$S:=\{0,8_e,12_{c-e=d  '}, 14_{c  '},15, 16_d***20_c \rightarrow\}$ has \
$H=\{c  '-1\}=\{13\}$, $t=0,\ \ \ell=3$,    $ \ \ e\neq
2\e+t$.}
\end{ex}

\begin{prop} \label{HH} Assume \ $\s<c  '+d  '-d$. \ Let $k:=min\{n\in\nat\ |\ d  '-n\notin S\},\ \
h:=d-c  '$, \ $s:= d  '+c  '-k-1$.  
We have  
\enu

\item   $s\leq 2d  '\II c  '-d  '\leq k+1\II d-d  '\leq k+h+1$.

\item  If $\ \s< d  '-k$   \ and    \ $\e\leq k+h +1$, \ then $\ \  s_m\geq s \geq c+d-e.$
    
\item   If   $\ 1\leq k<\e$,   $\ c  '-d  '\leq  k+1$ and   $\{d  '-\e,...,d  '\}\setminus\{d  '-k\}\subseteq
S$, \ then $\ c+d-e\leq s\leq s_m\leq 2d  '$.

\denu
\end{prop}
Proof. (1) is obvious by the assumptions.   \\
(2) We have $[d  '-k-\e+1, d  '-\e]\cap\nat\subseteq S$ (\ref{remacuti}.1). Further
   by (\ref{cof}.2) we have $\gamma(s)=-1$; in fact the assumption $\e\leq k+h +1$ implies
$c  '-d  '=d-d  '-h\leq \e-h\leq k+ 1$, and so $s\leq 2d  '$; since
$[s-d  ',d  ']\cap\nat\subseteq S, $ then $\gamma(s)=-1$  (\ref{cof}.2).
 Since    $\s<d  '$, then $ d-c  '\leq \e-2\ $, further $\e\leq k+h+1$;
therefore 
$$d  '-k-\e+1\leq s-d=d  '+c  '-k-1-d\leq d  '-\e.$$
 Hence $s-d\in S$. Moreover $s+1-c  '=d  '-k\notin S$. Then $\ s_m\geq s $ \ by Table \ref{corof} ($b)$.\\
To prove that $s\geq c+d-e$, recall that $c  '-d  '\leq k+1$. Then by assumption we have
 $\s<d  '-k<c  '-k-1\leq d  '$. \ Then $c  '-k-1\in S$ and by (\ref{rec0}.3) we get   $c  '-k-1 \geq c-e+\e$. \ Hence \\ 
 \centerline{   $s=d  '+c  '-k-1\geq d-\e+c  '-k-1\geq  c+d-e$ (\ref{cororecall}.3$a$)}
(3)  By   assumption  $\s<d  '-h\leq d  '$, and so $[d  '-h-\e, d  '-\e]\cap\nat\setminus\{d  '-k-\e\}\subseteq S$. Hence $[d  '-h-\e,d  '-k-1]\cap \nat\ \setminus \{d  '-k-\e\}\subseteq S$. Now recalling that $k<\e$ we get:
$$ d  '-h-\e\leq s-d<d  '-k-h\leq d  '-k.$$ 
Hence  \ $s-d\in S $: in fact  $s-d\neq d  '-k-\e$ because $s-d\geq d  '-k-\e+1.$ Further $s+1-c  '\notin S$ and $\gamma(s) =-1$ by (\ref{cof}.2) since $s\leq 2d  '$ and $\{s-d  ',...,d  '\} \subseteq S$.  Then $s\leq s_m$  \  by Table \ref{corof} ($b$). The inequality $s\geq c+d-e$ can be proved as in (2). For each element $2d  '<u< c  '+d  '$ we have $d  '\geq  u+1-c  '> s+1-c  '=d  '-k$; hence $u+1-c  '\in S  '$ then $\nu(u+1)\geq \nu (u) $ by Table \ref{corof} ($d)$.

\begin{coro} \label{H2} Suppose $\ \s<c  '+d  '-d$ and $\n H\leq 2$. Then \ $s_m\geq c+d-e$.

\end{coro}
Proof. If $\ \n H=0$, we have $s_m=\s+d$ and we are done by (\ref{H01}.1)
and (\ref{cong1}).
\\  If  ($\ \n H=2 $ \ and \ $H=\{d'+1,c'-1\}$), \ or \ $\n H=1$,    then \ either $\ s_m=\s+d$, or $\ s_m=2d  '$ (\ref{H01}.2); now  see (\ref{cong1}). \\ Finally assume that $H=\{d  '-k\}\cup
\{d  '+1\}$, with
$k\geq 1$.  
 In this case we have $c  '-d  '=2\leq k+1$, since $k\geq 1$.
 Hence the claim   $s_m\geq c+d-e$ \ follows by  (\ref{interv}), if
$k>\e$,  and by (\ref{HH}.3) if $k<\e$.
\quad $\diamond$\\

 \subsection{ Case $\ell=2 .$ }

If $\ \ell=2  $ ,  the conjecture (\ref{co}) is true, more precisely by {\rm \cite[Thm 5.5]{ot2} } we have: 

\begin{prop} \label{casol2}   Assume $\ \e=2$, then \ $s_m\geq c+d-e$ and \enu 
\item $s_m=\s+d $ \ if $\left[\begin{array}{ll}  
 
 t\leq 2, \\
 t=4   \ \ \\
  t\geq 5\ \ and\ \ d-3\in S.   \\

 \end{array}\right.$ 
\item  $s_m=2d-4 $ \  if  \ $ \left[\begin{array}{ll}
 either&   t=3 \ \  and \ \ d-6\notin S \\
 or&    t\geq 5\ \ and\ \ d-3\notin S. \end{array}\right.  $
\item  $s_m=2d-6 $ \  if  \   $t=3 $ \ and \ $d-6\in
S$\ \ $($all the remaining cases$)$.  
\denu
\end{prop}

Proof. The value of    $s_m$ is known by \cite[ Thm. 5.5 ] {ot2}. Another proof can be easily deduced by   \  (\ref{remacuti}.1), \  ( \ref{all}.2), \  (\ref{HH}.3), \  (\ref{interv1}.2), \  Table \ref{corof} $(d)$, \ (\ref{int0}. $\! e,  g$).
The inequalities   $s_m \geq  c+d -e$  now follow  respectively by  (\ref{cong1}) and by (\ref{interv}.3). \quad$\diamond$

\subsection{ Case $\ell=3 .$ }
If $\e=3$, we compute explicitly the possible values of $s_m$ and we show that the conjecture (\ref{co}) holds.
 
 \begin{notat}\label{set3} {\rm  (1) If $s_i=2d-k\in S $, $k\in
\nat$, let $$M(s_i ) \! :=  
 \{(s_h,s_j)\in S^2\ |\  s_i =s_h+s_j, s_h\leq d,s_j\leq d\}.$$
Note that    $M(s_i)=\{(d-x,d-y)\in S^2,\ |\  0\leq x,y\leq k,\ x+y=k\}$ and   that for $s_i\geq c$, we have  
$s_{i+1}-c=d-\e-k$;  \  for short it will be convenient to use the
following notation.
$$(*)\quad\left \{\begin{array}{lll}
\ \Sigma:= \{ z\in \nat,\ |\   z\leq d,\ d-z\in S\}  \vspace{0.2cm}\\
  (c,h)\in S\times \Sigma, \ h=d+c-s_i \  {\rm instead\ of\ the\ pair}\
 (c,s_{i}-c)\in N(s_i).  \end{array}\right.$$}
\end{notat}
If $\ell=3$, then $e\geq t+\e+1=t+4$  \  $(\ref{rec0}.2)$. To calculate the value of $s_m$,  we shall  assume 
$\s<c  '+d  '-d $, otherwise $s_m=\s+d$ by (\ref{all}.1). Then we have 
$t\geq 3, \  d-3\in S$ and $d-3\leq d  '$,   by (\ref{minor2d}.2) and by
(\ref{cororecall}.3$a$).  Three cases are possible:\\
$$\left.\begin{array}{lll} 
Case\ A:&S=\{0,e,...,  d-3,*,d-1,d,***,c=d+4\rightarrow\}& d  '=d-3\\
Case\ B:& S=\{0,e,..., d-3,*\ \ *\ ,d,***,c=d+4\rightarrow\}& d  '=d-3\\
Case\ C:& S=\{0,e,..., d-3,d-2,*,d,***,c=d+4\rightarrow\}& d  '=d-2.\\ 
\end{array}\right.$$
To describe $M(2d-k)$ we shall use the notations  $(*)$ fixed in (\ref{set3}) and
forwhen necessary for an element $2d-k$ we shall list all the  
 pairs $(x,y)\in M  '(2d-k)$   and the pair
$(c,\e+k+1)\in S\times \Sigma$  (the pairs underlined $\underline{(\ , \ )}$ not necessarily belong to $\Sigma^2$).
\begin{prop} \label{3} Assume $\e=3$. Then $s_m\geq c+d-e$. More precisely the values of $s_m$ can be computed as follows.
\end{prop}
{\bf Case A.} \ We have: \ \ 
  $ s_m=\left[\begin{array}{lll}
   \s+d&if   &either \ t\in [0,7]\setminus \{5\}\ \ or \ \ (t\geq 8, d-5\in S)\\
 2d-7&     if&   t\geq8,\ d-5\notin S \\ 
  
&     if&   t=5:\\
 2d-6&     if&     d-9\notin S \\
 2d-7&     if&  d-9\in S,d-10\notin S \\
 2d-9&  if &  \{d-9,d-10\}\subseteq S,\ d-12\notin S\\
 2d-10&  if & \{d-9,d-10,d-12\}\subseteq S     
  \end{array}\right.$  \spa\\
  Proof.
 $S=\{0,e,..., d-3,*,d-1,d,***,c=d+4\rightarrow\},\ \ $ with $e\geq \e+t+1= t+4\ (\ref{rec0}.1).$
\\
 First  we have $s_m=\s+d$ if $t\leq 2d-c  '-d  '=4$ and  $s_m<\s+d$,  if  $t=5$    by (\ref{all}.1 and 2). Hence we can assume $t\geq 5$, so that  $d-4=d-1-\e\in S,\ d-3-\ell=d-6\in S $, i.e.,   $\{0,1,3,4,6\}\subseteq 
\Sigma$. \par
 If $t=5$ we have $[d  '-\e,d  ']\cap \nat \subseteq S$ and
 $2d  ' =2d-6<\s+d=d  '+c  '-1$. \ \
  We obtain that  \ $2d-10\leq s_m\leq 2d  '$, \ $e\geq 2\e+t$   and $s_m\geq c+d-e$ by (\ref{interv}). 
More precisely we can verify that:\par
$s_m=
\left[\begin{array}{lll}
 2d-6&if \ 9\notin \Sigma& s_m=2d  ' \geq c+d-e\\
 2d-7&if \ 9\in \Sigma\ and \ 10\notin \Sigma& s_m=c+d-11\geq
c+d-e
\\ 2d-9&if\ 9\in \Sigma,\   10\in \Sigma, 12\notin \Sigma&s_m=c+d-13\geq
c+d-e\\
&&in\ fact \ 10\in\Sigma\I e\geq 14\\
2d-10& if \{9,10,12\}\subseteq\Sigma& s_m\geq c+d-e.
\end{array}\right.$\\
Note that in this case we have $\ d+d  '-\e-t+1=2d-2\e-t+1=2d-10\ $ and  this bound
  is achieved if $\{9,10,12\}\subseteq \Sigma$ (with $e\geq 16$).   See, e.g.       $S=\{0,16,34,36,37,39,40_{\s},41,   42   ,43_{d  '} ,  45, 46_{d}, 50_{c} \rightarrow\} $.
\par If $t\geq 6$ we have $\ \s\leq 2d  '-d $ \  and  we consider the following subcases. \par
If \ $t=6$, then $\ \s=2d  '-d=s_m$ \ by (\ref{all}.4).\par 
If \ $t\geq 7$ and $\ 5\in \Sigma$, \ one has $[d  '-\e,d  ']\cap\nat\subseteq
S$ and $\ \s<2d  '-d $, hence $\ s_m=\s+d$, by (\ref{interv}). \par 
 If 
$t\geq 7$ and $5\notin \Sigma$, we know that $s_m\leq 2d  '$ by
(\ref{all}.5); one can compute directly that\\ \centerline{ $\nu(2d  ' )<\nu(2d  '+1)$ 
(see Table \  \ref{corof} ($c$))       \ and that \ \ 
$\nu(2d-7)>\nu(2d-6)$,}\ hence $s_m=2d-7=2d  '-1.$ \ Since $d-6=d  '-\e\in
S, $ we get $e\geq 2\e+1+6=13$ (\ref{rec0}.3), \ and so $s_m\geq 
c+d-e+2.$\quad  \spa \\  
 {\bf Case B.}  \ \ We have: \ \ $ 
     \left[\begin{array}{lllll}
Case \ \  t\leq 3:&s_m   = \s+d \ \ \ \\
Case \ \  t\geq 4,\ d-5\notin S  : & s_m=2d-6\\
Case \ \  t\geq 4,\  d-5\in S ,  d-4\in S  : \\
 \hspace{0.4cm}\ if\ \ t\in\{4,5\}, &s_m\in[2d-9, 2d-6]&  \\
\hspace{0.4cm}\ if\ \ t\geq 6, &s_m \ =\s+d  \\
Case \ \  t\geq 5,\ d-5\in S,   \ d-4\notin S   :\\
\hspace{0.4cm}\ if\ \ t\in\{5,6,8\}, &s_m \ =\s+d  \\

\hspace{0.4cm}\ if\ \ t=7, &s_m \in\{2d-8, 2d-11\} \\

\hspace{0.4cm}\ if\ \ t\geq 9, \ \ d-7\notin S, &s_m =2d-8\\
\hspace{0.4cm}\ if\ \ t=9, \ \ d-7\in S,   &s_m \in\{2d-10,2d-11,2d-13\} \\

\hspace{0.4cm}\ if\ \ t\geq 10, \ d-7\in S,   &s_m =\s+d. \\

  \end{array}\right.$\vspace{0.2cm}\\
Proof.
 $S=\{0,e,..., d-3,*\ \ *\ ,d,***,c=d+4\rightarrow\},\ \  e\geq t+4$.\\
 As in case A we can see that   $s_m=\s+d$ \ if  \ $t\leq 3$ and $s_m< \s+d $ for $t=4$.
Suppose $t\geq 4$. Then  $
 \ \ \{ 0,3,6\}\subseteq \Sigma$.  
We deduce the statement by means of the  following table: \spa \\
$\left[\begin{array}{lllllll}

 2d-3&(0,3)  \\
 2d-4&\underline{(0,4)}  &\underline{(c,8)}&  \\

 2d- 5&\underline{(0,5)} &\underline{(c,9)} \\
  2d-6 &(0, 6) (3,3 )&\underline{(c,10 ).} \end{array}\right.$ \\
\spa  
 \par
 
If $\ 4\in \Sigma,\ 5\in \Sigma$, we have $[d  '-\e,d  ']\cap\nat \subseteq S$ and by 
applying (\ref{interv}) we get $s_m\geq c+d-e$. More  precisely, \ one
can easily verify that for
$t\in\{4,5\}$ we have $2d-9\leq s_m\leq 2d-6$, \  for $t\geq 6$ we have
$\s\leq 2d  '-d$, then by (\ref{remacuti}.1) and by (\ref{interv1}.2)
we get $\ s_m=\s+d$ .
 \par 
If $\ 5\notin   \Sigma $, we have $s_m=2d-6$.
 \par 
If $\ 4\notin   \Sigma, \  5\in   \Sigma  $:\\ 
 we have
$s_m=\left[\begin{array}{lllllll}2d-5 & \II &\ t=5\\ 2d-6 & \II &\ t=6 \ ( 8\in
\Sigma, \  9\notin \Sigma );
\end{array}\right.
\spa\\$ 
the remaining cases to consider satisfy
 $\{0,3, 5,6,8,9\}\subseteq \Sigma,4\notin \Sigma,$ with $t\geq 7$,  \ $s_m<2d-6$: \spa\\
 $\left[\begin{array}{lllllll} 
 2d-7 &  \underline{(0,7 ) }  &\underline{(c,11 )}  \\
2d-8 & (0,8) (3,5) &\underline{(c,12 )}\ \ s_m=2d-8\II
 \left[\begin{array}{lllllll} 7\notin \Sigma\ \  or \\ 7 \in \Sigma, 11\notin \Sigma \  (\I 7\leq t\leq 8)
\end{array}\right.\\ 
&otherwise \ \  7, 11\in \Sigma: & \{0,3, 5,6,7,8,9,11\}\subseteq \Sigma,4\notin \Sigma\\
 2d-9&(0,9)(3,6) &\underline{(c,13 )}\\
2d-10& \underline{(0,10)}(3,7) (5,5)&\underline{(c,14 )}\ \ s_m=2d-10\II
 10\in\Sigma, \ 13\notin\Sigma (\I 9\leq t\leq 10)\\& otherwise & 
\left[\begin{array}{lllllll} 

either& ( \alpha)\  \  10, 13\in \Sigma
\\ or& (\beta)\  \ 10\notin \Sigma \ (t=7)

\end{array}\right.  \end{array}\right.$\spa\par
\quad 
Case $( \alpha)$: \ $\{0,3, 5,6,7,8,9,10,11,13\}\subseteq \Sigma,4\notin \Sigma\ \
(\I t\geq 9)$:
\spa\\
  $\left[\begin{array}{lllllll} 
2d-10&(0,10)(3,7)(5,5) &\underline{(c,14 )}\\ 
2d-11&(0,11)(3,8)(5,6) &\underline{(c,15 )}\ \ s_m=2d-11\II 14\notin \Sigma\\
&&(\I t=9\ if \ 12\notin \Sigma,\  t=11\ if \ 12\in \Sigma)\\
& otherwise \ \  14\in \Sigma: &\{0,3,
5 \longleftrightarrow 11,13\}\subseteq
\Sigma,4\notin \Sigma\\ 2d-12&\underline{ (0,12)}(3,9)(5,7)(6,6)
&\underline{(c,16 )}\ \ s_m=2d-12\II t=12\\ &otherwise&
\left[\begin{array}{lllllll} either& (\alpha1)\ \ 12\notin \Sigma (\II t=9) \\
or& (\alpha 2)\ \ 12\in \Sigma,\ \ 15\in \Sigma
\end{array}\right.\\
2d-13&(0,13)(3,10)(5,8)(6,7) &\underline{(c,17)}\ \ s_m=2d-13 \left[\begin{array}{lllllll}   in \ case \
(\alpha1)\\ 
 in \ case \ (\alpha2) \II t=13

\end{array}\right.\\
& otherwise \ 16\in\Sigma &\{0,3,
5 \longleftrightarrow  16\}\subseteq
\Sigma\ 4\notin \Sigma:\\   2d-14&(0,14) (3,11)(5,9)(6,8)(7,7)
&\underline{(c,18)}\ \ s_m=2d-14 \II t=14\\ & otherwise \ 17\in\Sigma,
& \{0,3, 5\longleftrightarrow 15,16,17\}\subseteq
\Sigma,\ 4\notin \Sigma... \end{array}\right.$\spa\par

 Clearly  in cases $(\alpha2)$, for each $\ t\geq
13$ \  we get \ $s_m= \s+d.  \ $  \spa\par
\quad Case $(\beta)$: \ $\{0,3, 5,6,7,8,9, 11, \}\subseteq \Sigma,\
4,10\notin
\Sigma\ (t=7)$:
\spa\\
  $\left[\begin{array}{lllllll} 
2d-11&(0,11)(3,8)(5,6) &\underline{(c,15)}&s_m=2d-11.\\
 \end{array}\right.$\spa\\

 \noindent{\bf Case C}.  \ \  We have:  $ s_m=\left[\begin{array}{lll}
&     if&   t=3:\\
   2d-4&if   & d- 4\in S ,\ d-7\notin S \\
 2d-5&if   &  (\{d-4,d-7 \}\subseteq S ,\ d-8\notin S ) \ or\ (d-4\notin S)\\
 2d-7&     if&   \{d-4,d-7,d-8\}\subseteq S \\
  
&     if&   t\geq 4:\\
 \s+d&     if&    \ \ d-4\in S \\
 2d-5&     if& \ \ d-4\notin S. 

 \end{array}\right.$\spa\\
 Proof.
$ \ \ S=\{0,e,..., d-3,d-2,*,d,***,c=d+4\rightarrow\}, d  '=d-2  $.\\
As above we see that $t\geq 3$,\ \
$d-3,d-5\in S$ \  (i.e. $\{0,2,3,5\}\subseteq \Sigma$), $e\geq 7$.
 Consider the table:\spa\\
$\left[\begin{array}{lllllll} 2d-2&(0,2)& \\
 2d-3&(0,3) &\underline{(c,7)}\\
 2d-4&\underline{(0,4)}(2,2) &\underline{(c,8)}& s_m=&2d-4
&if\  4\in \Sigma, 7\notin \Sigma \\ 
2d- 5&(0,5) (2,3)&\underline{(c,9)}& s_m=&2d-5 &if
\left[\begin{array}{lllllll}   4\notin \Sigma \ or\\
 4,7\in \Sigma,  8\notin \Sigma. 

\end{array}\right.\\
if&\{4,7,8\}\subseteq \Sigma&&then& e\geq 12\ \ &(d-8+e\geq c)\\
 2d-6 &\underline{(0,6)} (2,4) (3,3 )&\underline{(c,10)}&\\
2d-7 &(0,7) (2,5)(3,4)  &\underline{(c,11)} 
\end{array}\right.$
\spa\par
\par If  $t=3$,
then $6\notin
\Sigma$:  we get $s_m=2d-7$.\par
If $t\geq 4$ and   $4\in \Sigma$,  we have $[d  '-\e,d  ']\cap\nat\subseteq S $
and $\s\leq 2d  '-d$.  Then $s_m=\s+d\geq c+d-e$ by (\ref{interv}).\par
If $4\notin S$, by the above table we deduce that $s_m=2d-5$. 
\quad  $\diamond $

\subsection{Semigroups with  CM type
$\ \tau \leq 7$.} As a consequence of the above results we obtain lower bounds or  the exact
value of $\ s_m$ for semigroups with small Cohen-Maculay type. First, in the next
lemma we collect  well-known or easy relations among   the CM type
$\tau$ of $S$ and the other  invariants.
\begin{lemma}\label{tipo} Let $\tau$ be the CM-type of the semigroup $S$  as
in {\rm(\ref{set1})}.\enu
\item $\n H+\e \leq \tau\leq e-1$ 
\item Assume
$\tau=\ell,$ then $H=\emptyset$ and the following conditions are
equivalent:
\enua 
\item $\ell=e-1$
\item $\tau=\ell,\ c  '=d.$
\item $d=c-e$.
\item $S=\{0,e,2e,...,ke\rightarrow\}$.
\denu

\item If $c  '>c-e$, then $\tau\geq \ell+1$ and $\tau=\ell+1\I H=\{c  '-1\}$.    

\item Assume $\s\leq d  '$ and $\tau=\ell+1$. Then $\left[\begin{array}{ll}
e\in\{2\ell+t-1,2\e+t\},&if \ \ \s=d  '\\e=2\ell+t,&if \ \ \s<d  '.\end{array}\right.$ 

\denu
\end{lemma}
Proof. (1) \ Clearly every gap $h\geq c-e$ belongs to $S(1)\setminus S$,
in  particular
$\{d+1,...,d+\ell\}\cup H\subseteq  S(1)\setminus S$. The inequality $\tau\leq e-1$. is
well-known.\\
(2) (3)  are   almost immediate.\\
 (4). We have $\n H\leq 1$ by (1), \  $\s-\e<\s\leq d  '<c  '-1$. 
If $\n H=0$, then $c'=c-e$ (\ref{H01}) and so $e=2\e+t$ (\ref{cororecall}.2). If $\n H=1 \I 
 \s-\e\notin H$ and $d'=c'-2$: it follows   $e\leq 2\ell+t$ by (\ref{rec0}.4). Now apply (\ref{rec0}.3) and (\ref{coroH}.2$c$): if $\ \s<d  '$, then $\s+1\in S$, hence 
$e\geq 2\ell+t$ and so $e=2\e+t$. Further $\ \s=d' \I c'\geq c-e+\e\I e\geq c-c'+\e=d+2\e-d'-1= 2\e+t-1$.
\begin{ex} {\rm \enu \item    We recall that in
general 
$\ c  '=c-e$ does not imply $\tau=\ell$. For instance, 
 let \\ $S=\{0,10_{e=d  '},16_{c-e},17,18,19,20,21,22,23,24_d,26_c\rightarrow\}$. Then
$\tau=5\neq \ell$. 
\item Analogously $H=\{c  '-1\}$ does not imply $\tau=\ell+1$:
 \\
$S=\{0,10_e,16_{c-e},17,18,19,20_{d  '}, 22_{c  '=d},26_c\rightarrow\}$ has
$\ell=3,\ \tau=5$. 
 \item  In (\ref{tipo}.4) the conditions    $e=2\ell+t$ and $\s<d'$ are not equivalent, further   when $\s=d'$ both the cases with $\ \tau=\e+1,\ \s=d'$ are possible.
For instance  \\
$ \{0,9_{e=c-e},10,11_{d  '},13_{c  '},14_d,18 _c\rightarrow\}$ \ has \ 
$t=\ell=3,\ e=2\ell+t,\ \s=d  ',\ \ \tau=\e+1$;\\
$ \{0,8_{e},9_{d  '},11_{c  '},12_d,16 _c\rightarrow\}$ \ has \ 
$t=\ell=3,\ e=2\ell+t-1,\ \s=d  ',\ \ \tau=\e+1$.\
\item There exist semigroups with $H=\{c  '-1\}, \ \s<d  ',\ \tau=\ell+1$ as in (\ref{tipo}.4):
\par $S=\{0 *...* 11_e* **15_{d-e}*** 19,20,21,22,23_{d  '}*25_{c  '}\ 26_d *** 30_c\rightarrow\} $,\\ has
 \ $\ell=3,\ t=5,\ e=2\ell+t,\ \tau=4$.

\denu}
\end{ex}
 Now we deduce bounds for $s_m$ when   $\ \tau\leq 7$.
     \begin{prop}   \label{tau}  For each $\tau\leq 7$ we have $s_m\geq c+d-e$.
 More precisely when $\s<c  '+d  '-d$  we have the following results.
  
  {\rm  
\enu 
\item   $\tau\leq 3$. We have: $s_m=\left[\begin{array}{llc}
 2d-4& {\rm if} \  S\ {\rm non\!-\!acute},\ \tau=t=3 \ (\ell=2)\\
\s+d&  {\rm in\ the\ other\ cases}
\end{array}\right.$\\
  \cite[5.9]{ot2} \cite[4.13]{ot1}

	 \item   $\tau=4$.  We have \  $\e\leq 4 $ and the following subcases.
\item[]
		 If $\e=4(=\tau)$, then $H=\emptyset  $ (\ref{tipo}.1), therefore $S$ is acute with
$\ s_m=\s+d$ \  (\ref{niedi}.4).\\ 
  If $\e\leq 3$ we are done by the previous (\ref{casol2}), (\ref{3}),
(\ref{niedi}.4) and (\ref{H01}) (recall $\ell=1\I   S$ is acute). More precisely
we get:\\
 $s_m=\left[\begin{array}{llc}
 2d-4& if  \left[\begin{array}{ll}
\ell=2\ and \ either \ (t=3,\ d-6\notin S)\ or\ (t\geq5,\ d-3\notin S)\\ 
\ell=t=3, e=9,\ c  '=d , \ d  '=c  '-2,\ d-4\in S \end{array}\right.\\ \vspace{-0.2cm}\\

2d-6& if  \left[\begin{array}{ll}
\ell=2,\  t=3,d-6\in S \\
\ell=3,\ t=5,\ e=11,\ c  '=d-1, d'=c'-2\end{array}\right.\\ \vspace{-0.2cm}\\
\s+d&  in\ the\ other\ cases.
\end{array}\right.$

\item   $\tau=5$. As above  we know $s_m$ in every case:

		\enua 
		\item  If $\ \e\leq 3$ we can deduce $s_m$ by  (\ref{casol2}), (\ref{3}),
		\item If $\ 4\leq \e\leq 5$, then    we are done by (\ref{H01}), since $\n
H\leq 1$. 
		\denu
				 \item   $\tau=6$.   We can calculate $s_m$  as follows:
		 
		\enua 
		\item If \  $\e\leq 3$ as in (3.$a$). 
		\item If \  $5\leq \e\leq 6$, then    we are done by (\ref{H01}), since $\n
H\leq 1$.
		\item If \  $\e=4$ and $H\subseteq \{c  '-2,c  '-1\}$, we have the value of $s_m$  by (\ref{H01}) and
(\ref{H01}.2).
		\item    If \  $\e=4$ and $H=\{d  '-k,c  '-1\}$, with $  k\geq 1$, then $d  '=c  '-2$, and the bounds for $s_m$  are given in (\ref{interv}) if $k>\e$, and in (\ref{H2})  \ if \ $\e> k\geq 1 $ (in fact $c  '-d  '=2\leq k+1$). 
		\denu
		
	 \item   $\tau=7$. We have $\e\leq 7$ and the following subcases.
\enua
\item  If \ $\e\leq 3$, then $s_m$ is known as in (3.$a$).
\item If \ $6\leq \e\leq 7$ then $\ \n H\leq 1$ and we are done by (\ref{H01}).
\item If \  $\e=5$, then $\ \n H\leq 2$ and we are done by (\ref{H01}), (\ref{H2}).
\item If \  $\e=4$, then $\ \n H\leq 3$ and we are done if $\ \n H\leq 2$ by (\ref{H01}), (\ref{H2}).\\ 
If \  $\e=4 , \ \   \n H= 3,$ \ consider   the following subcases
\enui 
\item   $H=[d  '+1,c  '-1]\cap \nat $: then $s_m$ is given in (\ref{H01}). 
\item $H=\{d  '-k,c  '-2,c  '-1\} ,\ k\geq 2 $. If $k<\e$, then $s_m$ is given in (\ref{HH}.3). \  If $k\geq\e$,  apply (\ref{interv}).
 \item $H=\{d  '-1,c  '-2,c  '-1\} $: this case cannot exist.  In fact since \\   $S=\{e,....,   d  '-2,*, d  ',*,*,c  ',...,d,c\rightarrow\} $ and by  the assumption   $\s<d  '$, we obtain    $ c  '-\e\in S$ and $c  '-\e=c  '-4=d  '-1\notin S$, \ impossible.
 \item  $H=\{d  '-j,d  '-k,c  '-1\}, \ \ j>  k\geq 1   $, hence \\
 $S=\{e,...,d  '-2,...,   d  ' ,*, c  '  , ...,d(\leq c  '+2),...,d+5\rightarrow \}$, with $2=c  '-d  '\leq  k+1$.  As in the proof of (\ref{HH}.2) for $s:= d  '-k+c  '-1$ we have $\gamma(s)=-1$, $s+1-c  '\notin S$, $s-d\in S\II s-d\neq d  '-j$.  Hence (Table \ref{corof} ($b$)) \ $s_m\geq s$ \ if $s-d\neq d  '-j$, i.e., $d  '-k+c  '-1-d\neq d  '-j$, i.e., $d-c  '\neq j-k-1$.\\  Four  subcases:
  $\left[\begin{array}{lll}
  j=k+1&\I& s_m\geq s \ \ if\ \ d\neq c  '\\
  j=k+2&\I& s_m\geq s \ \ if\ \  d\neq c  '+1\\
  j=k+3&\I& s_m\geq s \ \ if\ \  d\neq c  '+2\\
  j\geq k+4 & \I &s_m\geq s \\
  & ({\rm since \ }&   d-c  '\leq \e-2=2,\ j-k-1\geq 3).
  \end{array}
  \right.$\\
  In the remaining three situations  we can see that   $\ [d  '-\e,d  ']\cap\nat\subseteq S$, therefore   $s_m $ is given by (\ref{interv}):

   -\ If $j=k+1$ and $\ d=c  ' $,  since $\s<d  ',\ \e=4$ we get $\{   d  '-4, d-4=d  '-2,   d  '\}\subseteq S$. Since there are two consecutive holes, then   $k\geq 5$. It follows $[d  '-\e,d  ']\cap\nat\subseteq S$.\\
   - \ If $j=k+2$, \  and $ \ d=c  '+1$, we have $c  '-4=d  '-2,\ d  '-1=d-4$ and   $\s< d  '-1$ (\ref{HH}.1) . Therefore \  $S=\{e,...,d  '-5, d  '-4,d'-3,*, d  '-2,d  '-1 ,    d  ',  *, d  '+2=c  ',d  '+3=d,  d+5\rightarrow \}$. We deduce $[d  '-\e,d  ']\cap\nat\subseteq S$.\\
    - \ If $j=k+3$, and \ $d=c'+2$, \ analogously we deduce $[d  '-\e,d  ']\cap\nat\subseteq S$.\quad$\diamond$
 \denu 
 \denu 
\denu}
\end{prop}

\subsection{The value of s$_m$ for semigroups of multipicity $e\leq 8$.}
\begin{coro}\label{eminore8} For each semigroup $S$ of multiplicity $e\leq 8$ we have $s_m\geq c+d-e$.
\end{coro}
Proof. Since $\tau\leq e-1$ the result follows by (\ref{tau}).
 \quad $\diamond$\spa\par
\subsection {Almost arithmetic sequences and Suzuki curves.}
Recall that a semigroup $S$ is   {\it generated by an almost arithmetic sequence}  (shortly AAS)  if  \\ \centerline{$S=<m_0, m_1,..., m_{p+1}\ ,n> $}
  with \ \ $ m_0\geq 2,\ \ m_i=m_0+\rho\ i,\quad \forall\ i=1,...,p+1,  \ \  and  \  GCD(\rho, m_0,n)=1 . $  (The  embedding dimension of $S$ is $embdim \ S=p+2$).
  
   \begin{prop} \label{AAS} Let  $S$ be an AAS semigroup of    $embedding\  dimension$ \  $\mu$; \ then  $\ \tau\leq 2(\mu-2)$. 
\end{prop}
Proof. It is a consequence of \cite[3.3\ -\ 4.6\ -\ 4.7\ -\ 5.6\ -\ 5.7\ -\ 5.8\ -\ 5.9]{ps} after
suitable   calculations.
\begin{coro}  If  $S$ is an AAS semigroup with $ embdim \ S\leq 5$ then $s_m\geq c+d-e$.
\end{coro}

Proof. It is an immediate consequence of (\ref{tau}) and (\ref{AAS}). \quad$\diamond$\\

\noindent As another corollary we obtain the value of $s_m$   for the Weierstrass semigroup of a
$Suzuki\ curve$, that  is a plane curve $C$ defined by the equation $$y^b-y=x^{a}(x^b-x),\ {\rm with}
\  \ a=2^n,\ b=2^{2n+1},\ n>0.$$
Some applications of these curves to AG codes can be found for example in  \cite{m}. 
\begin{prop} \label{suzu}
  If  $S$ is the Weierstrass semigroup  of a Suzuki Curve, then $S$ is symmetric, therefore    $s_m=\s+d$. \end{prop}
Proof. In \cite[Lemma 3.1]{m} is proved that the Weierstrass semigroup  $S$ at a rational place of the function field of $C$ is  generated as follows: 
 $$S=<b,b+a,b+\displaystyle{\frac{b}{a}},1+b+\displaystyle{\frac{b}{a}}> $$
We have $b=2a^2$, \  with \ $a=2^n$, \ hence $S=< 2a^2,\ 2a^2+a,\  2a^2+2a,\ 2a^2+2a+1>$
Then consider the semigroup $$S=< 2a^2,\ 2a^2+a,\  2a^2+2a,\ 2a^2+2a+1>, \  \ a\in \nat.$$  \\
If $a =1, $ then $S=<2,3>$.\\
If $a>1$, then $S$ is  generated by an almost arithmetic sequence, and    \ $embdim(S)=4$; \ in fact \  $$S=<m_0,m_1,m_2,n>,\ \ {\rm with}\ m_0=2a^2,\ m_1=m_0+a,\ m_2=m_1+a,\ n=m_2+1 .$$   Since $S$ is  $AAS$, we shall compute the Apery set ${\cal A}$   by means of 
    the algorithm described in \cite{ps}: \\ 
    let  
$p=embdim(S)-2=2$ and for each
$t\in
\nat$,\\ let
$\ q_t,r_t\ $ be the (uniqe) integers such that  $\ t=pq_t+r_t(=2q_t+r_t), \ q_t\in\integ,\ r_t\in\{1,2\}$, \\ 
let
$g_t=q_tm_2+m_{r_t}$, \ i.e., \  $g_t= 
\left[\begin{array}{ll}
(q_t +1)m_2& if \ r_t=2\\
\ q_tm_2+m_1& if \ r_t=1
\end{array}\right.$\ \ 
(in particular $g_0=0$). \\
Then by   \cite{ps} the Apery set ${\cal A}$ of $S$ is:
 $\{\ g_t+hn\ \ | \ \ 0\leq t \leq 2a-1,\ 0\leq h\leq a-1 \ \}:$ 
therefore the elements of the Apery set are the $2a^2$ entries of the following matrix 
$$\left[\begin{array}{ccccccccccccc}
0&g_1&g_2&g_3&.&.&.&  g_{2a-1}\\
& {||}& {||}& {||}&.&.&.& {||}\\
&m_1&m_2&m_1+m_2&.&.&.&(a-1)m_2+m_1\\
n&g_1+n\ &g_2+\ n&.&.&.&.&.\\
2n&g_1+2n&g_2+2n&.&.&.&.&.\\
.&.&.&.&.&.&.&.&\\
.&.&.&.&.&.&.&.&\\
(a-1)n&g_1+(a-1)n&g_2+(a-1)n&.&.&.&.& g_{2a-1}+(a-1)n&\\
 \end{array}\right]$$
\indent Recall that a semigroup $S$ of multiplicity $e$ and Apery set   ${\cal A}$ is\par
$symmetric\II$ for each $s_i\in {\cal A},\ 0<s_i\neq s_e:=max\ \! {\cal A} $, there exists $s_j\in {\cal
A}$ such that
$s_i+s_j=s_e$.\\ In our case c   this condition is satisfied: in
 fact   $s_i= \left[\begin{array}{crl}
 \alpha m_2+h\ n,&  h\leq a-1,\ \ \alpha\geq 1&(1)\ \ or\\
 \alpha \ m_2+m_1+h\ n&\ \    0\leq \alpha,\  h\leq a-1&(2)\\
\end{array}\right.$ \\   further $s_e= (a-1)(m_2+n)+m_1$, 
and so  \\ \centerline{$s_e-s_i=
\left[\begin{array}{rll}
 (a-1-\alpha )m_2+m_1+(a-1-h)\ n  \in {\cal A}  &(1)\ \ or\\
 (a-1-\alpha) \ m_2+ (a-1-h)\ n  \in {\cal A} &(2)\\
\end{array}\right.$}
Since a  semigroup $S$ is symmetric if and only if  its   CM-type is one, then $\ s_m=\s+d\ $ by
(\ref{tau}.1).\quad$\diamond$

\end{document}